\documentclass[amssymb,prb,twocolumn,showpacs]{revtex4}
\usepackage{epsfig}
\usepackage{dcolumn}
\usepackage{amsmath}
\begin{document}
%\documentclass[12 pt,a4paper]{article} %selecciona el tipo de documento
%\usepackage[english]{babel} %selecciona el idioma
%\frenchspacing %trata los espacios despues de los puntos igual que los otros
%\usepackage{epsfig}
%\usepackage{amsmath}
%\usepackage[a4paper,dvips]{geometry}
%\geometry{textwidth=16 cm, textheight=22 cm}
%\begin{document}
\title{\bf\ Giant off-resonance  resistance  spike related phenomena
in irradiated ultraclean two-dimensional electron systems.}
\author{J. I\~narrea}
\affiliation{Escuela Polit\'ecnica
Superior, Universidad Carlos III, Leganes, Madrid, 28911, Spain}
\date{\today}
%%%%%%%%%%%%%%%%%%%%%%%%%%%%%%%%%%%%%%%%%%%%%%%%%%%%%%%%%%%%%%%%%%%%%%%%%%%%%%
%\section{Abstract}
\begin{abstract}
We report on theoretical studies of a recently discovered strong
radiation-induced magnetoresistance spike obtained in ultraclean two-dimensional
electron systems at low
temperatures. The most striking feature of this spike is that it shows up
on the  second harmonic of the cyclotron resonance and with an amplitude
that can reach an order of magnitude
larger than the radiation-induced resistance oscillations.
We apply  the radiation-driven electron orbits model
in the ultraclean scenario. Accordingly, we calculate the
elastic scattering rate (charged impurity) which will define the unexpected resonance spike position.
We also obtain the inelastic scattering rate (phonon damping),
that will be responsible of the large spike amplitude.
We present a microscopical model to explain the dependence of
the Landau level  width on the magnetic field for ultraclean samples.
We find that this dependence explains the experimental shift
 of the resistance oscillations with respect to the magnetic field found in this kind of  samples.
We study also recent results on the influence of an in-plane magnetic
field on the spike. We are able to reconcile the obtained different experimental response of both spike
and resistance oscillations versus an increasing in-plane field. The same model on the
variation of the LL width, allows us
to explain such surprising results based in the increasing disorder in the sample
caused by the in-planed magnetic field.
Calculated results are in good agreement with experiments.
These results would be of special interest in nanophotonics;
they could lead to the design of novel ultrasensitive microwave detectors.
\end{abstract}
%%%%%%%%%%%%%%%%%%%%%%%%%%%%%%%%%%%%%%%%%%%%%%%%%%%%%%%%%%%%%%%%%%%%%%%%%%%%%%
\maketitle
\section{ Introduction}
Quantum Hall effect and radiation-matter coupling\cite{ina1} are two
of the most remarkable
topics in Condensed Matter Physics. Accordingly, transport excited by radiation in a two-dimensional electron system (2DES) has been always
 a central topic in basic and applied research.
In the last decade it was discovered that when a 2DES in a perpendicular magnetic field ($B$)
%(a 2DES
%with a uniform and perpendicular magnetic field ($B$))
is irradiated, mainly with microwaves (MW), some striking effects are revealed:
radiation-induced  magnetoresistance ($R_{xx}$) oscillations  and zero
resistance states (ZRS) \cite{mani1,zudov1}.
These
remarkable effects show up at low $B$ and high mobility samples.
It is important to achieve a complete understanding of the physical mechanisms being
responsible of them and not only for the basic knowledge purpose but
also for the potential applications in nanoelectronics.
Different
theories have been proposed to explain these  effects
\cite{ina2,ina20,girvin,dietel,lei,ryzhii,rivera,shi,vavilov} but the physical
origin is still being questioned.
In the same way, a great effort has also been  made from the
experimental side
%, growing better samples, adding new features and
%different probes to the basic experimental setup, etc.
\cite{mani2,mani3,willett,mani4,smet,yuan,kons1,vk}.
Of course the  obtained experimental results always mean a real challenge
for the existent theoretical models. Therefore, a comparison of
experiment with theory could help to identify the importance of the
different approaches in these theories.
Thus, as an example,
it has been recently published experimental results on the
dependence of the oscillations with radiation power\cite{mani6,wiedmann}
where a very solid result has been obtained in terms of a sublinear relation, similar
to a square root.
Yet, some theories predicted a linear dependence\cite{dmitriev} between
$R_{xx}$ oscillations and radiation power.

One of the most interesting and challenging experimental
result, recently obtained\cite{yanhua,hatke2} and as
intriguing as ZRS,
consists in a strong resistance spike which shows up far off-resonance.
It occurs  at twice the
cyclotron frequency, $w\approx2w_{c}$\cite{yanhua,hatke2},
were $w$ is the radiation
frequency and $w_{c}$ the cyclotron frequency.
The amplitude of such a spike is
very large reaching an order of magnitude regarding the
radiation-induced  $R_{xx}$  oscillations.
In the same experiments it is also reported the behavior of the
resistance spike with different radiation frequencies, temperature ($T$) and
radiation power ($P$).
Remarkably, the only different feature in these
experiments\cite{yanhua,hatke2} is the use of ultraclean samples
with mobility $\mu\sim 3\times 10^{7}cm^{2}/V s$ and lower temperatures, $T\sim 0.4 K$.
Yet, for the previous, $"standard"$, experiments \cite{mani1,zudov1} the
mobility
is lower, ($\mu < 10^{7}cm^{2}/V s$) and  $T$ higher, ($T\geq 1.0 K$).
Much more recently experimental results on this off-resonance giant resistance
spike have been presented by the same authors\cite{yanhua2}. This time they
add an in-plane magnetic field to the standard experimental set-up. Remarkably, they
obtain that the radiation-induced resistance oscillations and spike present
different response to an increasing in-plane magnetic field.
Resistance oscillations are shifted to higher perpendicular magnetic field ($B$), when the
in-plane is increased. Yet, the spike approximately keeps the same position. Then,
the authors suggest that both effect are ruled by different physical
mechanisms.

Other strong resistance spikes in radiation-induced $R_{xx}$ oscillations experiments were previously
obtained around the cyclotron resonance $w\approx  w_{c}$\cite{smet} for
higher frequencies and power and being explained by resonant
heating of electrons.
Off-resonance phenomena are always remarkable phenomena in
all branches of Physics. They are counterintuitive with respect to
the basic resonance process, meaning a challenge to
the available theory.
Then,  it would be
important to fully understand these phenomena, not
only from the basic standpoint but also from the application side.
For instance, they could lead to the design and development of novel {\it ultrasensitive}
photon detectors\cite{manifoton} in the microwave and terahertz bands, where the technology is not
well-developed yet.

\begin{figure}
\centering \epsfxsize=3.9in \epsfysize=3.3in
\epsffile{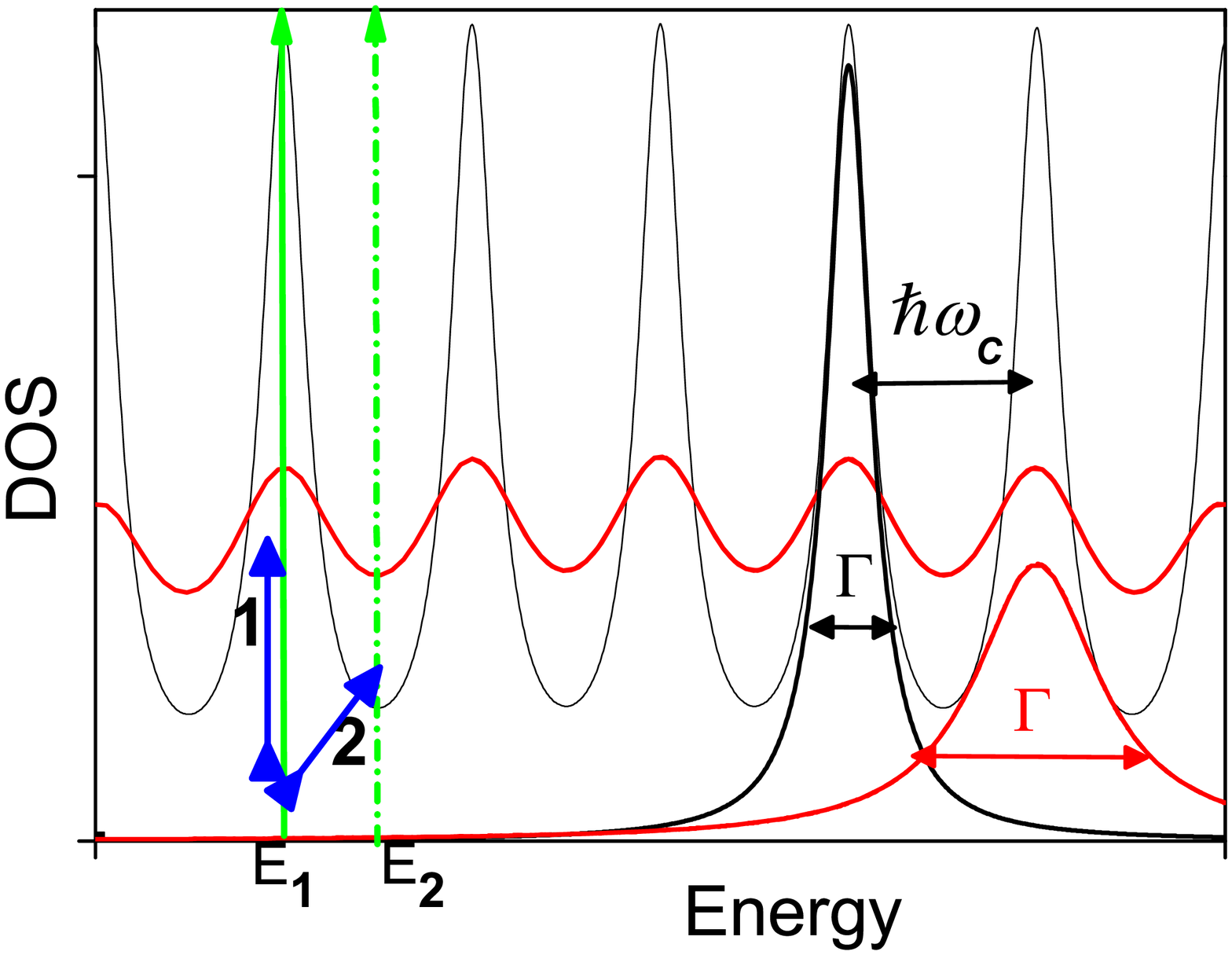}
\centering \epsfxsize=3.9in \epsfysize=2.0in
\epsffile{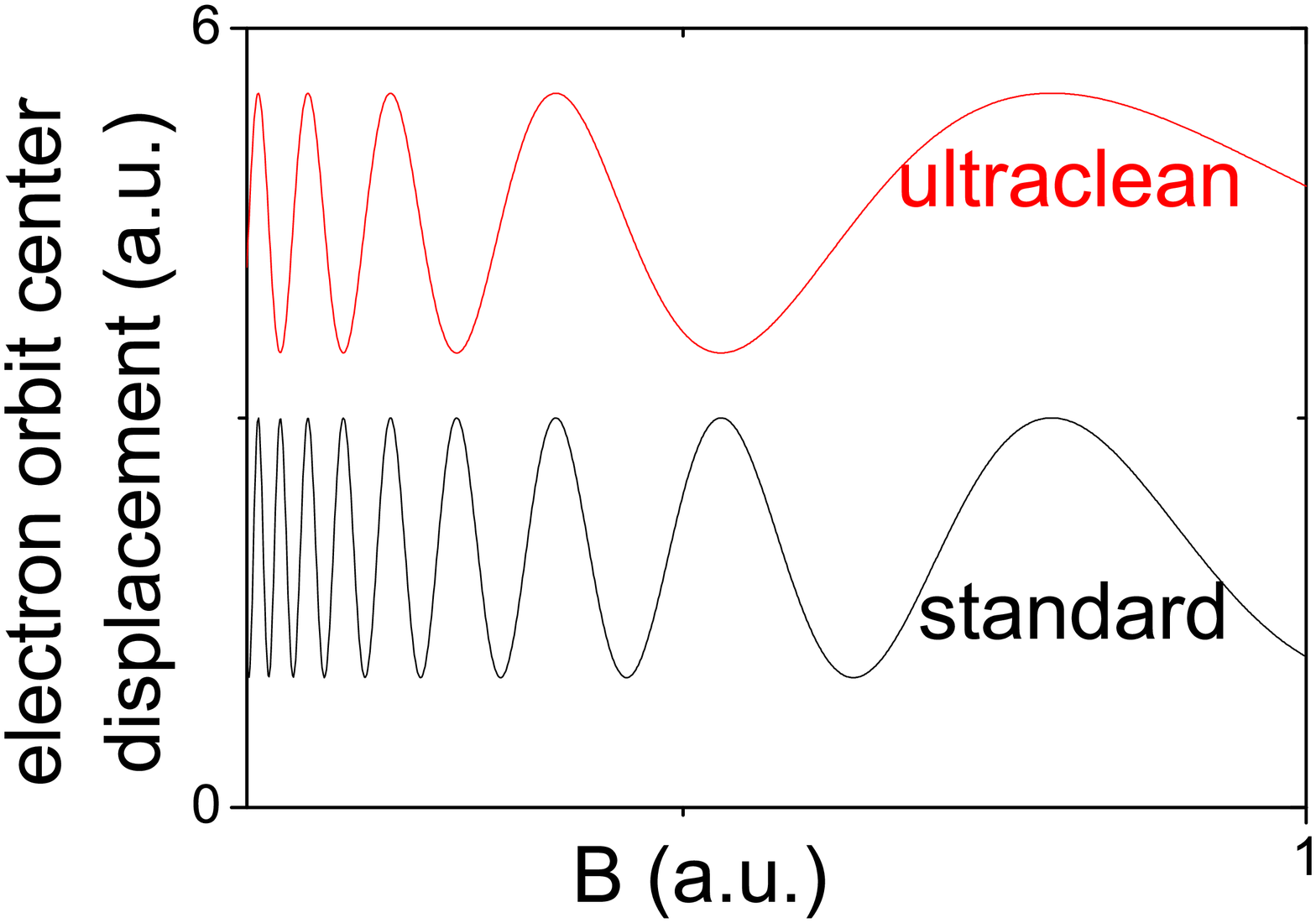}
\caption{
Upper panel: schematic diagram showing the density of Landau states simulated by Lorentzian functions
for wide and narrow Landau levels, where the  width of the states is indicated by  $\Gamma$. The arrow 1 corresponds to an
elastic scattering process and the arrow 2 to an inelastic one.
Lower panel:  schematic diagram showing the radiation-driven electron orbit center displacement
(two dimensional electron system) for
ultraclean and standard samples. We observe that the first one is
delayed regarding the latter as being driven by radiation of smaller frequency.}
\end{figure}

In this article, we
theoretically study this radiation-induced
 $R_{xx}$ spike applying
the theory developed by the authors, {\it the radiation-driven electron
orbits model}\cite{ina2,ina20,ina3,kerner,park}. According to it,
when a Hall bar is illuminated, the electron
orbit centers perform a classical trajectory consisting in a harmonic
displacement along the direction of the current at the $w$ frequency. This
motion is damped by the interaction of electrons
with the lattice ions and with the consequent  emission of acoustic phonons.
Thus, the  2DES moves
periodically at the radiation frequency altering dramatically the scattering
conditions and giving rise eventually to   $R_{xx}$ oscillations and ZRS.
We extend this model to a ultraclean sample,
obtaining that
all the scattering conditions are modified. Mainly because the LL, which in
principle are broadened by scattering,  become
very narrow in this kind of samples. This implies an increasing number of states
at the center of the LL sharing a similar energy. In between LL, it happens the opposite,
the density of states decreases dramatically  (see upper panel of Fig.1).
%\begin{figure}
%\centering \epsfxsize=3.0in \epsfysize=2.0in
%\epsffile{spikes01.eps}
%\epsfig{file=conducfig1.eps,angle=0,width=0.4\textwidth}
%\caption{
%}
%\end{figure}
 Therefore, the elastic scattering rate, due to charged impurities,
increases, meanwhile the inelastic one (phonon emission damping) decreases.
Thus, the elastic scattering time (inverse of elastic scattering rate)
decreases reaching a
limiting value  of half of the time of a standard sample.
This specific value is related  with the LL degeneracy.
For the irradiated electrons, half of the scattering time is physically equivalent as being
driven by radiation of frequency $w/2$. Accordingly, the
cyclotron resonance is shifted to a new
$B$-position around $w\approx2w_{c}$.
On the other hand, the inelastic scattering decreases and the emission of acoustic phonons is less efficient
producing a {\it bottleneck effect} which prevents from releasing the absorbed energy
 to the lattice. Finally the corresponding amplitude abruptly increases
giving rise to a strong resistance spike.

We also present a microscopical theoretical approach to  the dependence of
the LL width on the magnetic field for the regime of ultraclean samples.
We apply this approach in the framework of the general theory of the radiation-driven
electronic orbits model.
We find that this dependence is very important to explain the experimental shift
found in the resistance oscillations with respect to the magnetic field for this kind of  samples\cite{yanhua}.
We study also very recent results on the influence of an in-plane magnetic
field on the spike and the corresponding connection with
the resistance oscillations. We consider that the main effect of
adding this parallel magnetic field is to increase the disorder
perceived by the electrons in the sample. Within our model and among
the different sources of scattering,
inelastic scattering of electrons by interaction with acoustic phonons (emission)
is the most directly affected. The basic result is a
progressive damping of the spike and the whole radiation resistance response. This is observed
in experiments\cite{yanhua2}.
The total scattering rate reflects also this increasing disorder which broadens the LL;
now the LL width depends   on both the perpendicular and in-plane magnetic fields.
As feedback effect, this will eventually affect the charged impurity scattering rate.
This feedback and the LL degeneracy cut-off value explain the
different behavior of the spike and resistance oscillations with
the in-plane magnetic field.
Initially, all spike-related phenomena
were tried to be explained\cite{yanhua} as unique events
with an origin different  from radiation-induced  $R_{xx}$ oscillations.
Yet, in this article we consider  that the resistance spike and everything related with it, are the outcome
 of an extreme scenario of the radiation-induced oscillations in ultraclean samples and
therefore, sharing the same physics.

\section{ Theoretical Model}

\subsection{Summary of the radiation-driven electron orbit model}

The {\it radiation-driven electron orbits model}, was developed to explain
the $R_{xx}$ response of an irradiated 2DEG at low magnetic field. We first obtain
an exact expression of the electronic wave function.
Then, the total hamiltonian $H$ can be written as\cite{ina2}:
\begin{eqnarray}
H&=&\frac{P_{x}^{2}}{2m^{*}}+\frac{1}{2}m^{*}w_{c}^{2}(x-X)^{2}-eE_{dc}X +\nonumber \\
 & &+\frac{1}{2}m^{*}\frac{E_{dc}^{2}}{B^{2}}\nonumber-eE_{0}\cos wt (x-X) -\nonumber \\
 & &-eE_{0}\cos wt X \nonumber\\
 &=&H_{1}-eE_{0}\cos wt X
\end{eqnarray}
 $X$ is the center of the orbit for the electron spiral motion:
\begin{equation}
X=\frac{\hbar k_{y}}{eB}- \frac{eE_{dc}}{m^{*}w_{c}^{2}}
\end{equation}
$E_{0}$ the intensity for the MW field and $E_{dc}$ is the DC
electric field in the $x$ direction. $H_{1}$ is the hamiltonian
corresponding to a forced harmonic oscillator whose orbit is
centered at $X$. $H_{1}$ can be solved exactly \cite{kerner,park},
and using this result allows an exact solution for the electronic wave function\cite{ina2}:
\begin{widetext}
\begin{equation}
\Psi(x,t)=\phi_{n}(x-X-x_{cl}(t),t)
 exp \left[i\frac{m^{*}}{\hbar}\frac{dx_{cl}(t)}{dt}x+
\frac{i}{\hbar}\int_{0}^{t} {\it L} dt'\right]
\sum_{m=-\infty}^{\infty} J_{m}\left[\frac{eE_{0}}{\hbar}
\left(\frac{1}{w}+\frac{w}{\sqrt{(w_{c}^{2}-w^{2})^{2}+\gamma^{4}}}\right)\right]
e^{imwt}
\end{equation}
\end{widetext}
and the main result that we want to  point out is that
%\begin{eqnarray}
$\Psi_{N}(x,t)\propto\phi_{n}(x-X-x_{cl}(t),t)$
%\end{eqnarray}
where $\phi_{n}$ is the solution for the
Schr\"{o}dinger equation of the unforced quantum harmonic
oscillator. $J_{m}$ are Bessel functions and $L$ is a classical Lagrangian of
an electron in the presence of magnetic field.
%\begin{widetext}
\begin{figure}
\centering \epsfxsize=3.8in \epsfysize=3.45in
\epsffile{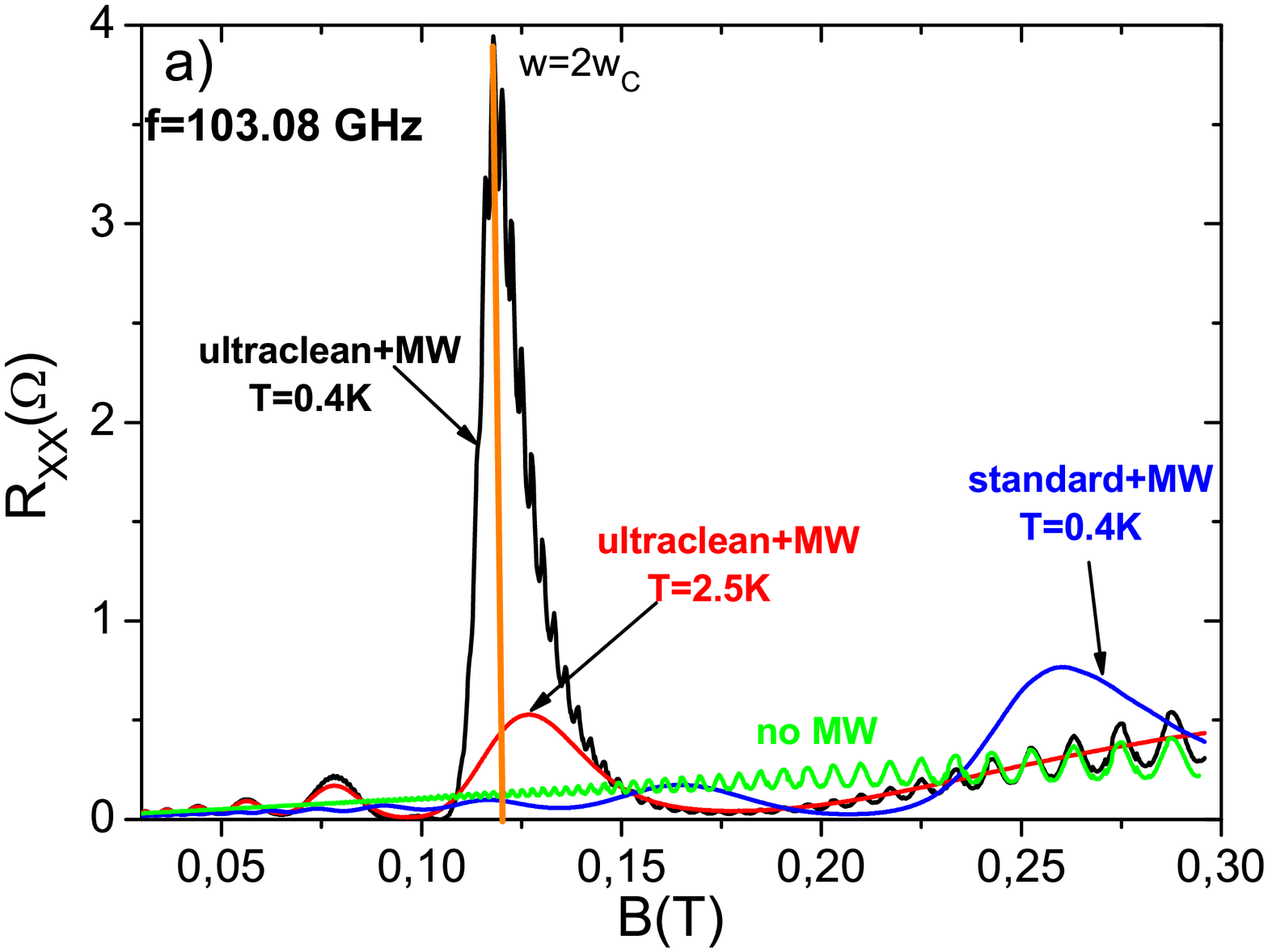}
\centering \epsfxsize=3.8in \epsfysize=2.5in
\epsffile{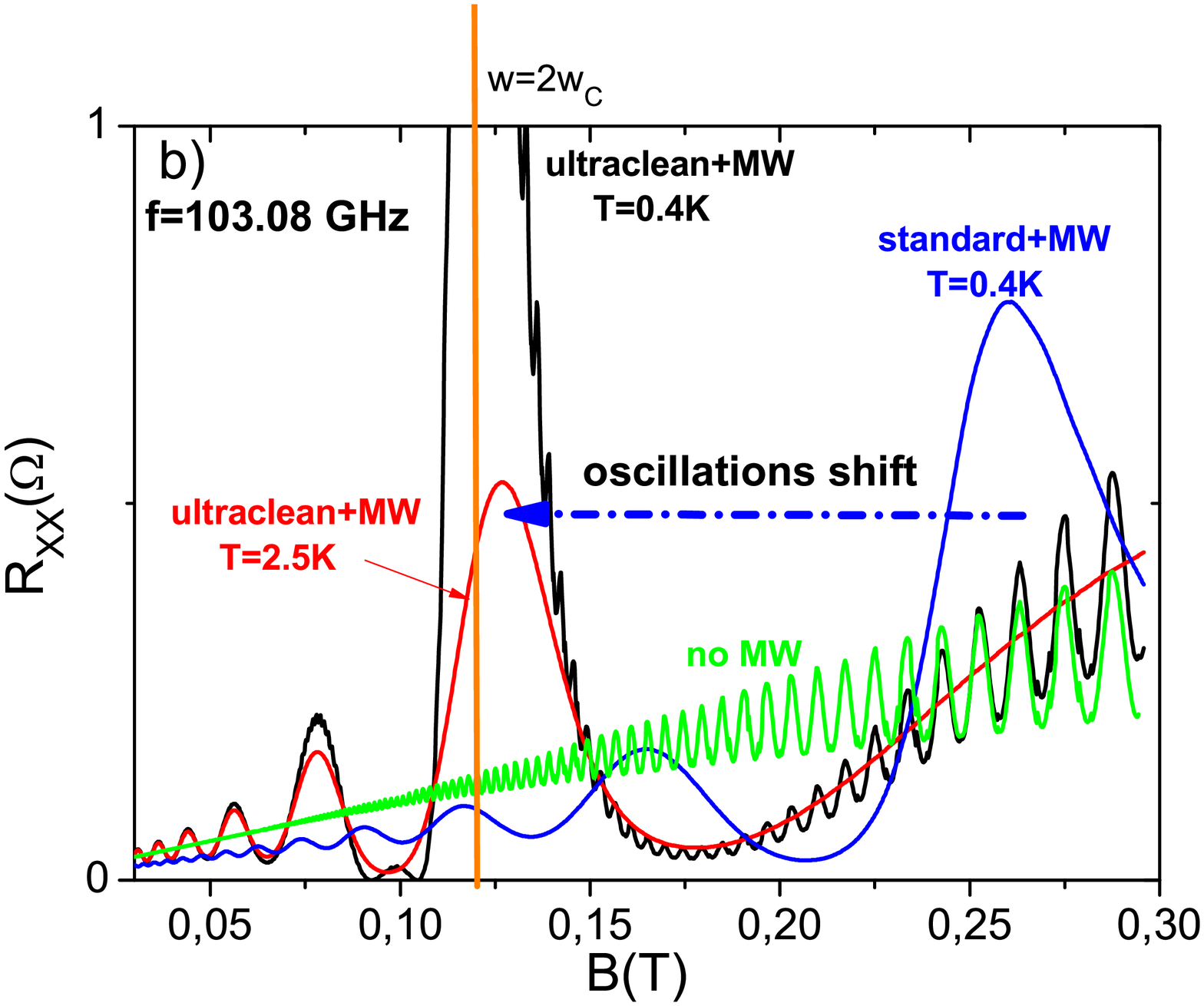}
\caption{a) Calculated irradiated  magnetoresistance vs static magnetic field for a radiation frequency of
$f=103.08 GHz$, different samples (standard and ultraclean) and temperatures. For
ultraclean sample and a temperature of, $T=0.4 K$, we observe a intense spike at $w\approx2 w_{c}$. Yet, for
higher temperature, $T\simeq 2.5K$, the spike vanishes.
For a standard sample  and low temperature, we obtain a curve with the usual
radiation-induced resistance oscillations  and zero resistance states
but the spike does not show up.
b) Blown-up of upper panel for lower magnetoresistance values.  We observe that  in ultraclean samples, radiation-induced oscillations are
shifted to lower magnetic field positions, compared to standard sample case.}
\end{figure}
%\end{widetext}
$x_{cl}(t)$ is the classical solution of a forced  and damped harmonic
oscillator:
\begin{eqnarray}
x_{cl}&=&\frac{e E_{o}}{m^{*}\sqrt{(w_{c}^{2}-w^{2})^{2}+\gamma^{4}}}\cos wt\nonumber\\
&=& A \cos wt
\end{eqnarray}
where $E_{0}$ is the MW electric field. $\gamma$ is a  damping factor
for the electronic interaction with the lattice ions giving rise to emission of acoustic phonons.
\begin{figure}
\centering\epsfxsize=3.7in \epsfysize=3.4in
\epsffile{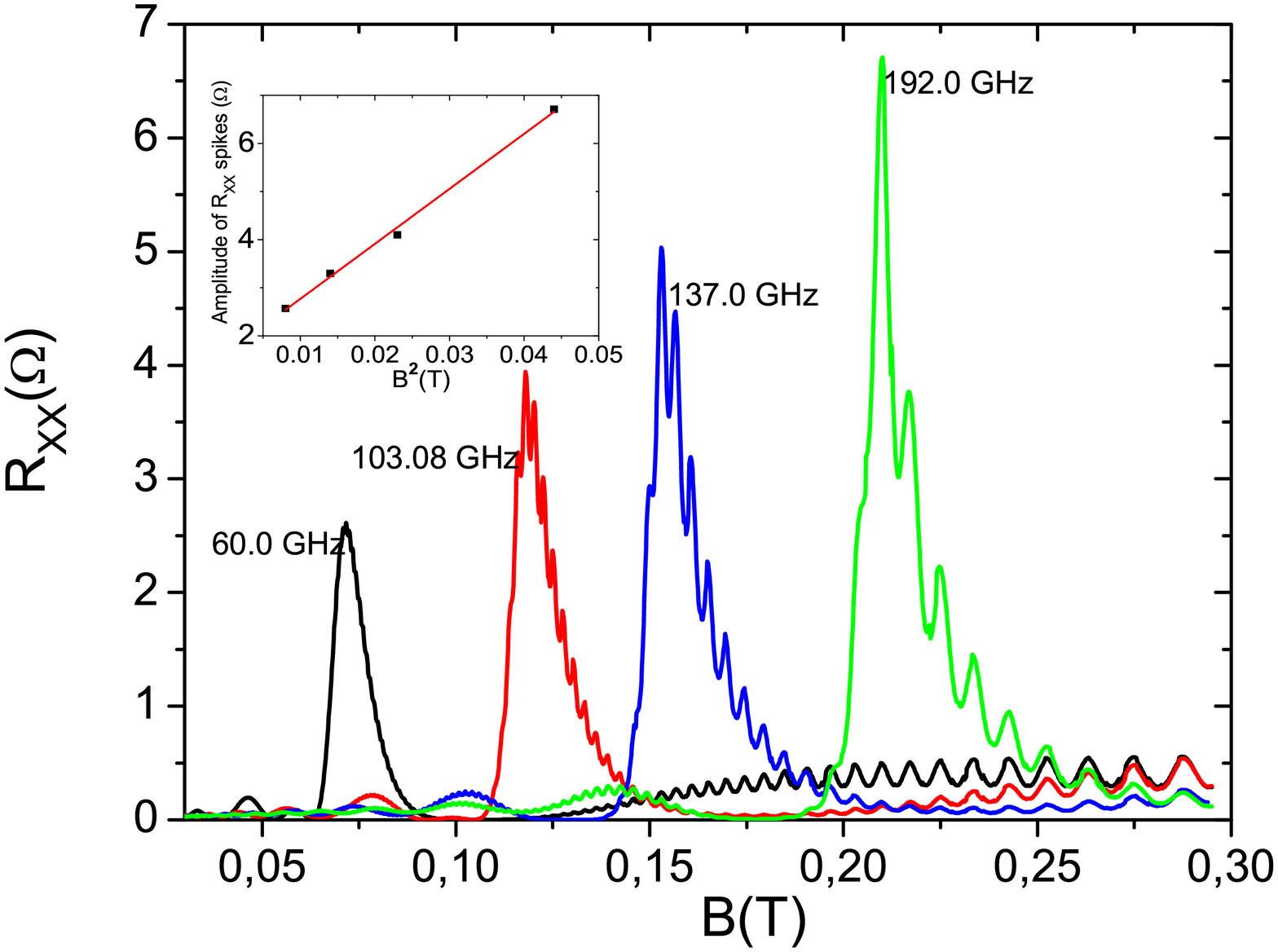}
\caption{Calculated irradiated magnetoresistance vs magnetic field for radiation frequencies $f=60.0, 103.08,
137.0$ and $192$ GHz. For each frequency we obtain a spike at $w\approx2w_{c}$.
The spike amplitude increases with the radiation frequency, and with  the
magnetic field as $B^{2}$. This dependence
can be shown in the inset, where we present the amplitude of the spike vs
the square of the magnetic field.}
\end{figure}
Then, the obtained wave
function is the same as the one of the standard quantum harmonic oscillator where the
center is displaced by $x_{cl}(t)$.
This implies that the electron orbit centers
oscillate harmonically at $w$.
 This $radiation-driven$ behavior will affect dramatically the
charged impurity scattering and eventually the conductivity.

Next,
 we calculate the scattering suffered by the electrons due to
charged impurities (elastic) applying time dependent first order perturbation theory. Thus, we
calculate the scattering rate\cite{ina2,ina4,ridley}
between two $oscillating$ Landau states (LS), the initial,  $n$,  and the
final, $m$:
\begin{equation}
W_{n,m}=\lim_{\alpha\rightarrow 0} \frac{d}{d t} \left|
\frac{1}{i \hbar} \int_{-\infty}^{t^{'}}<\Psi_{m}(x,t) |V_{s}|\Psi_{n}(x,t)>e^{\alpha t}d t\right|^{2}
\end{equation}
where $V_{s}$ is the scattering potential for charged impurities\cite{ando},
\begin{equation}
V_{s}= \sum_{q}\frac{e^{2}}{2 S \epsilon (q+q_{0})} \cdot e^{i
\overrightarrow{q}\cdot\overrightarrow{r}}
\end{equation}
$S$ being the surface of the sample, $\epsilon$ the GaAs dielectric
constant, and $q_{0}$ is the Thomas-Fermi screening
constant\cite{ando,davies}. After some  algebra we arrive at an intermediate
expression for the charged impurities scattering rate:
\begin{eqnarray}
W_{m,n}&=&\frac{2\pi}{\hbar}|<\phi_{m}|V_{s}|\phi_{n}>|^{2}\nonumber\\
&&\times\frac{2eB}{h}S\sum_{m=0}^{\infty}\frac{1}{\pi}\left[\frac{\Gamma}
{(E_{n}-E_{m})^{2}+\Gamma^{2}}\right]
\end{eqnarray}
$\Gamma$ is the LL width and
the last part of the expression represents the  density of final Landau states, $\rho_{m}$,  in the
elastic scattering jump,
\begin{equation}
\rho_{m}=\frac{2eB}{h}S\sum_{m=0}^{\infty}\frac{1}{\pi}\frac{\Gamma}{(E_{n}-E_{m})^{2}+\Gamma^{2}}
\end{equation}
\\
$E_{n}$ and $E_{m}$ are the corresponding LL energies for the initial
and final states respectively.
% for the initial state "n" and for the final "m" respectively.

Once we know the  scattering rate, we consider that  when
an electron undergoes a scattering process jumping from the initial state
to the final one, it advances an average effective distance\cite{ina2,ina3,ridley}, $\Delta X^{rad}$,
given by
\begin{equation}
\Delta X^{rad}= \Delta X^{0}+A\cos w\tau
\end{equation}
 where $\Delta X^{0}$ is the
effective distance advanced when there is no radiation field present.
Then, following Ridley\cite{ridley} we can obtain the expression
for the longitudinal conductivity $\sigma_{xx}$ according
to the expression,
\begin{equation}
\sigma_{xx}\propto \int dE \frac{\Delta X^{rad}}{\tau_{uclean}}
\end{equation}
being $E$
the energy.
Then, we get to the final expression
which reads,

\large
\begin{widetext}
\begin{equation}
\sigma_{xx}=\frac{e^{7}m^{*2}B n_{i}S}{\pi\epsilon^{2}\hbar
 ^{6}q_{0}}\left[\Delta X^{0}+ A\cos
w\langle\tau\rangle\right]^{2}
\left[1+2e^{\frac{-\pi\Gamma}{\hbar w_{c}}}+e^{\frac{-\pi\Gamma}
{\hbar w_{c}}}\frac{X_{S}}{\sinh X_{S}}\left(\cos\frac{2\pi E_{F}}{\hbar w_{c}}\right)\right]
\end{equation}
\end{widetext}
\normalsize
where $n_{i}$ is the impurities density and  $X_{S}=\frac{2\pi^{2}k_{B}T}{\hbar w_{c}}$.
To obtain $R_{xx}$ we use
the relation
$R_{xx}=\frac{\sigma_{xx}}{\sigma_{xx}^{2}+\sigma_{xy}^{2}}
\simeq\frac{\sigma_{xx}}{\sigma_{xy}^{2}}$, where
$\sigma_{xy}\simeq\frac{n_{i}e}{B}$ and $\sigma_{xx}\ll\sigma_{xy}$.
Therefore, we reach  an expression
for $R_{xx}$ :
\begin{equation}
\large
R_{xx}\propto\frac{e
E_{o}}{m^{*}\sqrt{(w_{c}^{2}-w^{2})^{2}+\gamma^{4}}}\cos
w\tau
\end{equation}

The radiation power, $P$, can
be related with $A$ through the well-known formula that gives
radiation intensity $I$ (power divided by surface) in terms of the radiation electric field
$E_{0}$: $I=\frac{1}{2}c\epsilon_{0}E_{0}^{2}$, where $c$ is the
speed of light in vacuum and  $\epsilon_{0}$ is the  permittivity in
vacuum. If we want to express only the power in terms of the radiation
electric field we have to take into account the sample surface. In the particular case of GaAs
 we can readily obtain:
\begin{equation}
\large
 P = \frac{1}{2} c_{GaAs} \epsilon \epsilon_{0}E_{0}^{2} S
\end{equation}
where $c_{GaAs}$ is the speed of light in GaAs and $\epsilon$
is the GaAs dielectric constant.  Accordingly,
\begin{equation}
E_{0}\propto
\sqrt{P}
\end{equation}
 Then, substituting in the expression of $A$, we
obtain that $R_{xx}$ varies with $P$ following an square root law:
\begin{equation}
\large
R_{xx}\propto \frac{e\sqrt{P}
}{m^{*}\sqrt{(w_{c}^{2}-\frac{w}{2}^{2})^{2}+\gamma^{4}}}\cos \frac{w}{2} \tau
\end{equation}
Thus, we expect that the $R_{xx}$ response will grow according to the square root of $P$, i.e., following a
sublinear dependence which is finally reflected in the amplitude, $A\propto P^{0.5}$.

\subsection{Far off-resonance position of the giant resistance spike}

The precise shape of broadened LL in real systems (with disorder) remains, even currently, controversial.
Usual  assumptions are Gaussian\cite{coler,rai,wang}, Lorentzian\cite{coler2,itsko,drago, hiro}
and semielliptic\cite{ando}, being the two first, the most commonly used. For instance, some published
experimental results\cite{potts,zhou} indicate that they could be equally described or explained assuming  either
a Lorentzian or Gaussian profiles. More recently, other works
related with radiation-induced $R_{xx}$ oscillations and ZRS, have
used Lorentzian shapes for LL in the corresponding theoretical models\cite{studen,yanhua}
In our case, dealing with ultraclean samples (very narrow LL), and low $B$ and $T$, there is no much theoretical or experimental information
about the real profile of the broadened Landau states. We have to consider that  the samples used in the experiments\cite{yanhua, hatke2}
with extremely high mobilities have been very recently  obtained. Then,
we are facing a completely brand new scenario requiring more theoretical and
experimental work. Therefore, as a first approach and
following T. Ihn\cite{ihn}, we have assumed for the broadened density of Landau states
a Lorentzian function being the width independent of the LL index.

If we rewrite eq. (8) considering that  $E_{m}= \hbar w_{c}(m+\frac{1}{2})$,
\begin{equation}
\rho_{m}=\frac{2eB}{h}S\sum_{m=0}^{\infty}\frac{1}{\pi}\frac{\Gamma}{(E_{n}-\hbar w_{c}(m+\frac{1}{2}))^{2}+\Gamma^{2}}
\end{equation}
Now, this last expression can be further developed if we apply the Poison sum rules\cite{ihn},
\begin{widetext}
\large
\begin{equation}
\sum^{\infty}_{m=0}f(m+\frac{1}{2})=\int_{0}^{\infty}f(x)dx+2\sum_{s=1}^{\infty}(-1)^{s}\int_{0}^{\infty}f(x)\cos(2\pi xs)dx
\end{equation}
\end{widetext}
and then, we can reach for $\rho_{m}$,
\begin{equation}
\rho_{m}=\frac{m^{*}}{\pi \hbar^{2}}\Bigg\{1+2\sum_{s=1}^{\infty}(-1)^{s} \cos\left[ \frac{2\pi s E_{n}}{\hbar w_{c}} \right]
exp\left[-\frac{\pi\Gamma s}{\hbar w_{c}}\right] \Bigg\}
\end{equation}
that with $E_{n}= \hbar w_{c}(n+\frac{1}{2})$
we
%\begin{equation}
%\rho_{m}=\frac{m^{*}}{\pi \hbar^{2}}\Bigg\{1+2\sum_{s=1}^{\infty}(-1)^{s} \cos\left[ \pi s (2n+1) \right]
%exp\left[-\frac{\pi\Gamma s}{\hbar w_{c}}\right] \Bigg\}
%\end{equation}
finally obtain,
\begin{equation}
\rho_{m}=\frac{m^{*}}{\pi \hbar^{2}}\Bigg\{1+2\sum_{s=1}^{\infty}exp\left[-\frac{\pi\Gamma s}{\hbar w_{c}}\right] \Bigg\}
\end{equation}

Normally in this calculations  if the exponent is not very small   it is enough
to consider  the first term $s=1$.  Yet, since we are dealing with high mobility samples and very small
$\Gamma$ the latter approximation can not be applied. Then if the total sum is carried out we obtain\cite{grads}:
\begin{equation}
\large
\sum_{s=1}^{\infty}exp\left[-\frac{\pi\Gamma s}{\hbar w_{c}}\right] =\frac{exp\left[-\frac{\pi\Gamma}{\hbar w_{c}}\right]}{1-exp\left[-\frac{\pi\Gamma}{\hbar w_{c}}\right]}
\end{equation}
that translated into  the final density of states,
\begin{widetext}
\begin{equation}
\large
\rho_{m}=\frac{m^{*}}{\pi \hbar^{2}}\Bigg\{1+2\frac{exp\left[-\frac{\pi\Gamma}{\hbar w_{c}}\right]}{1-exp\left[-\frac{\pi\Gamma}{\hbar w_{c}}\right]} \Bigg\} \\
=\frac{m^{*}}{\pi \hbar^{2}}\Bigg\{\frac{1+exp\left[-\frac{\pi\Gamma}{\hbar w_{c}}\right]}{1-exp\left[-\frac{\pi\Gamma}{\hbar w_{c}}\right]} \Bigg\}
\end{equation}
\end{widetext}
Substituting this result in eq. (7), we arrive at the final expression for the charged
impurities scattering rate which reads:
\begin{equation}
\large
W_{m,n}=W_{0}\left( \frac{1+exp\left[\frac{-\pi\Gamma}{\hbar w_{c}}\right]}{1-exp\left[\frac{-\pi\Gamma}{\hbar w_{c}}\right]}\right)
\end{equation}
where  $W_{0}$\cite{inaw} represents the charged impurities
scattering rate for a $standard$ mobility sample which is
recovered for wide $\Gamma$,
\begin{equation}
W_{0}=\frac{e^{5}B m^{*}n_{i}S}{\hbar^{4}\epsilon^{2}q_{0}^{2}}
\end{equation}

As we said above,  for ultraclean samples $\Gamma$ is very small and
for experimental magnetic fields\cite{yanhua,hatke2}, it turns out that
$\Gamma<< \hbar w_{c}$. Then, if $B$ increases, the exponent $\frac{-\pi\Gamma}{\hbar w_{c}}$
decreases making $W_{m,n}$ continuously to increase.
However, there exists a cut-off defined by the LL degeneracy:
\begin{equation}
LLdegen.=\frac{2eB}{h}
\end{equation}
In a regime of narrow LL the density of states $\rho_{m}$ can be approximated considering
that  in  an elastic scattering, such as the one caused by charged impurities, most processes take place at the center of the LL, sharing approximately the
same energy (see Fig.1 upper panel) and then $E_{m}\simeq E_{n}$.
This implies that,
\begin{equation}
\frac{2eB}{h}\frac{1}{\pi}\frac{\Gamma}{(E_{m}-E_{n})^{2}+\Gamma^{2}}
\rightarrow \frac{2eB}{h}\frac{1}{\pi\Gamma}
\end{equation}
In this regime of narrow $\Gamma$, in an interval
energy of $\hbar w_{c}$ we must have, in average, a number of states given by the
LL degeneracy and therefore:

\begin{equation}
\frac{2eB}{h}\frac{1}{\pi\Gamma}\times
\hbar w_{c}\rightarrow \frac{2eB}{h}
\end{equation}
\\

This is only fulfilled if,
\begin{equation}
\frac{\hbar
w_{c}}{\pi\Gamma}\rightarrow 1
\end{equation}
which  is a remarkable result
because translated into the expression  of eq. (22) we
obtain:
\begin{widetext}
\large
\begin{equation}
\frac{\hbar w_{c}}{\pi\Gamma}\rightarrow 1 \Rightarrow \left[
\frac{1+e^{\frac{-\pi\Gamma}{\hbar
w_{c}}}}{1-e^{\frac{-\pi\Gamma}{\hbar w_{c}}}}\right]\rightarrow 2
\Rightarrow W_{m,n}\simeq 2 \times W_{0}
\end{equation}
\end{widetext}
Then, the charged impurities scattering rate
increases with $B$ till a cut-off value of twice the
corresponding of a standard sample.
In terms of charged impurities scattering
time,
\begin{equation}
\tau_{uclean}=\frac{1}{2}\tau
\end{equation}
where
$\tau=1/W_{0}$ is the time of a standard sample and $\tau_{uclean}$ corresponds to a
ultraclean one.
Accordingly, the  ultraclean  scattering time,
%\begin{equation}
$\tau_{uclean}=\frac{1}{W_{mn}}$
%\end{equation}
turns out to be twice shorter
 than the standard and then, the scattering event is twice faster.

Within the framework of our theory\cite{ina2},
this implies that during the time $\tau_{uclean}$ and compared to
the standard for
the same $B$, the ultraclean
2DES appears to be displaced by radiation a smaller distance.
In other words, in terms of scattering, the
{\it ultraclean} harmonic motion (electron orbit center
displacement) is perceived as
delayed regarding the {\it standard}, as if electrons
were driven by radiation of smaller frequency (see Fig. 1 lower panel), more precisely, of half
frequency. Then, the radiation electric field, $E_{w}$, is perceived as,
\begin{equation}
E_{w}=E_{0}\cos \frac{w}{2}t
\end{equation}
by the ultraclean 2DES.
%This is
%physically equivalent to a larger scattering time ($\tau=2\tau_{uc}$), but
%with Landau states being driven by MW of smaller frequency
%($\frac{w}{2}$).
The conclusion is that in ultraclean samples,  during scattering, electrons $"feel"$ radiation
with half of the real frequency.
Applying
next the theory\cite{ina2}, we reach  an expression
for $R_{xx}$ in the ultraclean case:
\begin{equation}
\large
R_{xx}\propto\frac{e
E_{o}}{m^{*}\sqrt{(w_{c}^{2}-(\frac{w}{2})^{2})^{2}+\gamma^{4}}}\cos
\frac{w}{2}\tau
\end{equation}
According to it, now the resonance in $R_{xx}$ will take place at
$w\approx2 w_{c}$, as experimentally obtained\cite{yanhua,hatke2}.

\subsection{Intensity of the resistance spike}

The intensity of the $R_{xx}$ spike will depend on the relative value of
the frequency term, ($w_{c}^{2}-(\frac{w}{2})^{2}$), and the damping parameter $\gamma$ in
the denominator of the latter $R_{xx}$ expression.
When $\gamma$ leads the denominator the spike is smeared out. Yet,
in situations where $\gamma$ is smaller than the frequency term, the
resonance effect will be more visible and the spike will show up.
As we explained above, the parameter $\gamma$ represents the interaction of electrons with
the lattice ions, damping the electronic orbits motion and releasing
radiation energy in form of acoustic phonons. Therefore, $\gamma$ is given by\cite{ando}:
\begin{equation}
\gamma=\frac{1}{\tau_{ac}}\propto T \times\frac{2eB}{h}\sum_{m=0}^{\infty}\frac{1}{\pi}\left[\frac{\Gamma}
{(E_{n}-\hbar w_{ac}-E_{m})^{2}+\Gamma^{2}}\right]
\end{equation}
where $ w_{ac}$ is the frequency of acoustic phonons for experimental parameters\cite{yanhua,hatke2}
and the last term represents the density of final Landau states.
Following similar steps as before  we
obtain the expression:
\large
\begin{eqnarray}
\gamma&=&\frac{2\Xi^{2}m^{*}k_{B}T}{v_{s}^{2}\rho \pi \hbar^{3}<z>}\times \nonumber\\
&&\Bigg\{1+2\sum_{s=1}^{\infty}exp\left[-\frac{\pi\Gamma s}{\hbar w_{c}}\right]\cos\left[\frac{2\pi s \hbar w_{ac}}{\hbar w_{c}} \right]\Bigg\}\nonumber\\
\end{eqnarray}
\normalsize
where $\Xi$ is the acoustic deformation potential, $\rho$ the
mass density, $k_{B} $ the Boltzman constant, $T$ the temperature, $v{s}$ the sound velocity and $<z>$ is the effective
layer thickness. When the total sum inside brackets is carried out we obtain\cite{grads},

\large
\begin{eqnarray}
&&\sum_{s=1}^{\infty}exp\left[-\frac{\pi\Gamma s}{\hbar w_{c}}\right]\cos\left[\frac{2\pi s \hbar w_{ac}}{\hbar w_{c}} \right]=\nonumber\\
&&\nonumber\\
\nonumber\\
&&\large=\frac{e^{\left[-\frac{\pi\Gamma }{\hbar w_{c}}\right]}  \big\{  \cos \left[\frac{2\pi  \hbar w_{ac}}{\hbar w_{c}}\right]- e^{\left[-\frac{\pi\Gamma }{\hbar w_{c}}\right] }\big\}}
{1-2e^{\left[-\frac{\pi\Gamma }{\hbar w_{c}}\right]}   \cos \left[\frac{2\pi  \hbar w_{ac}}{\hbar w_{c}}\right]+e^{\left[-\frac{2\pi\Gamma }{\hbar w_{c}}\right]} }\nonumber\\
\end{eqnarray}
\normalsize
Then, substituting in $\gamma$ we finally obtain,
\Large
\begin{widetext}
\LARGE
\begin{equation}
\LARGE
\gamma=\frac{2\Xi^{2}m^{*}k_{B}T}{v_{s}^{2}\rho \pi \hbar^{3}<z>}\large\Bigg\{ \frac{1-e^{\left[-\frac{2\pi\Gamma }{\hbar w_{c}}\right]} } {1-2e^{\left[-\frac{\pi\Gamma }{\hbar w_{c}}\right]}   \cos \left[\frac{2\pi  \hbar w_{ac}}{\hbar w_{c}}\right]+e^{\left[-\frac{2\pi\Gamma }{\hbar w_{c}} \right]} }  \Bigg\}
\\
\end{equation}
\end{widetext}
\normalsize

 In order to have an approximate but  more compact version of eq. (27) we can consider average values for
the  cyclotron energy and for the acoustic phonon energy in the case of experimental parameters.
 Thus, for the magnetic fields swept in the experiments we can take in average that the cyclotron energy is  $\hbar w_{c} \sim 2 \times 10^{-4}$ meV,
 and the phonon acoustic energy is, $\hbar w_{ac}\sim 1 \times 10^{-4}$ meV.
 Then, for those values the cosine term,
 \begin{equation}
\cos \left[\frac{2\pi  \hbar w_{ac}}{\hbar w_{c}}\right]\sim-1
\end{equation}
that gives an expression for the damping parameter which reads

\begin{equation}
\large
\gamma=\frac{2\Xi^{2}m^{*}k_{B}T}{v_{s}^{2}\rho \pi \hbar^{3}<z>}\left( \frac{1-e^{\frac{-\pi\Gamma}{\hbar w_{c}}}}{1+e^{\frac{-\pi\Gamma}{\hbar w_{c}}}}\right)
\end{equation}
%where we have considered that in average,  for narrow $\Gamma$ and experimental $B$, $\Gamma<\hbar w_{phon}< \hbar w_{c}$.

According to this last expression, for increasing $B$ the term inside brackets decreases and as a result, the
parameter $\gamma$ will get smaller, making increasingly difficult
the damping by acoustic phonon emission and the release of the absorbed energy to the lattice.
Therefore, we have
a {\it bottleneck effect} for the emission of acoustic phonons. Now it is possible
to reach a situation where $(w_{c}^{2}-(\frac{w}{2})^{2})^{2}\gtrsim\gamma^{4}$ making visible a resonance effect and, therefore,  giving rise
to a strong resonance peak at $w\approx2 w_{c}$.
For GaAs and standard experimental parameters\cite{mani1,zudov1}, we obtain that $\gamma\simeq 7-10 \times
10^{11} s^{-1}$ and that in a ultra-clean regime\cite{yanhua,hatke2} it decreases till $\sim 3.5\times 10^{11}$ .
This fixes a lower cut-off value for the radiation frequency where the resistance spike
could be observed. According to our calculations this would be
around $f=w/2\pi \approx 40-45 GHz$. Experiments have also pointed out
that at lower frequencies is much more difficult to observe resistance spikes.

\subsection{Variation of $\Gamma$ with the magnetic field
for ultraclean samples}

One important issue which plays a key role in the model presented above is the
dependence of $\Gamma$ on the magnetic field. Previous experimental and theoretical
works\cite{coler,rai,wang,coler2,itsko,drago, hiro} report that for lower $B$
(but higher than $1$ T), $\Gamma$ varies with
the magnetic field as $\sim\sqrt{B}$. Yet, for increasing
$B$, $\Gamma$ becomes independent.
For our brand new scenario of extremely high mobility
samples at very low $B$,  ($\sim 0.1$ T), and very low temperature,
($\sim 0.4$ K),
there is no much information, experimental or theoretical, about the actual
dependence of $\Gamma$ on $B$. Therefore, and as a first approach,
we have developed a simple microscopical description, based in scattering parameters,
of this dependence which starts from
previous  analytical expressions\cite{ihn,coler,wang,potts,sayaka,ando}.
Thus, following T. Ihn\cite{ihn}, we begin with the expression which relates
$\Gamma$ and $B$,
\begin{equation}
\large
\Gamma(B)=\sqrt{\frac{\hbar^{2}w_{c}}{2\pi \tau_{0}}}=\sqrt{\frac{\hbar^{2} e}{2\pi m^{*}\tau_{0}}}\sqrt{B}
\end{equation}
where $\tau_{0}$ is the zero magnetic field quantum scattering lifetime.
%This expression coincides o is very similar to previous expressions
%obtained by other authors\cite{coler,wang,potts,ando}.

For the experimental spikes\cite{yanhua,hatke2}, the
magnetic field used is very low, around $B\sim0.1 T$. Then, for
this regime we can assume that the variation of $\Gamma$ with $B$
is closer to a straight line than to a square root and proceed to
linearize the latter expression. Thus, we expand the
square root by a Taylor series around $B=0.1$ T, till
the first derivative term. Then,
 we can write,
\begin{equation}
\large
\Gamma(B)=\Gamma(0.1)+(B-0.1)\Gamma'(0.1)
\end{equation}
and finally obtain the expression which gives the
dependence of $\Gamma$ on $B$ for very low $B$
\begin{equation}
\large
\Gamma(B)\propto \frac{1}{2\sqrt{0.1}}\sqrt{\frac{\hbar^{2} e}{2\pi m^{*}\tau_{0}}} B^{p}
\end{equation}
where $p\rightarrow 1$. p is a phenomenologically introduced
parameter which takes into account that the dependence of $\Gamma$
on $B$ does not follow strictly a straight line but it is close
to it. In our case we have assigned $p\sim 0.90$.
%\begin{equation}
%$\Gamma(B)\propto B^{0.95}$.
%\end{equation}
This expression implies that, for extremely high mobility samples and very
low $B$, the Landau states
broaden almost linearly with the magnetic field affecting
all scattering processes such as elastic impurity scattering
 and inelastic phonon scattering. This dependence of $\Gamma$
 on $B$  has been used
throughout the above theoretical model on spikes.

\begin{figure}
\centering \epsfxsize=3.9in \epsfysize=4in
\epsffile{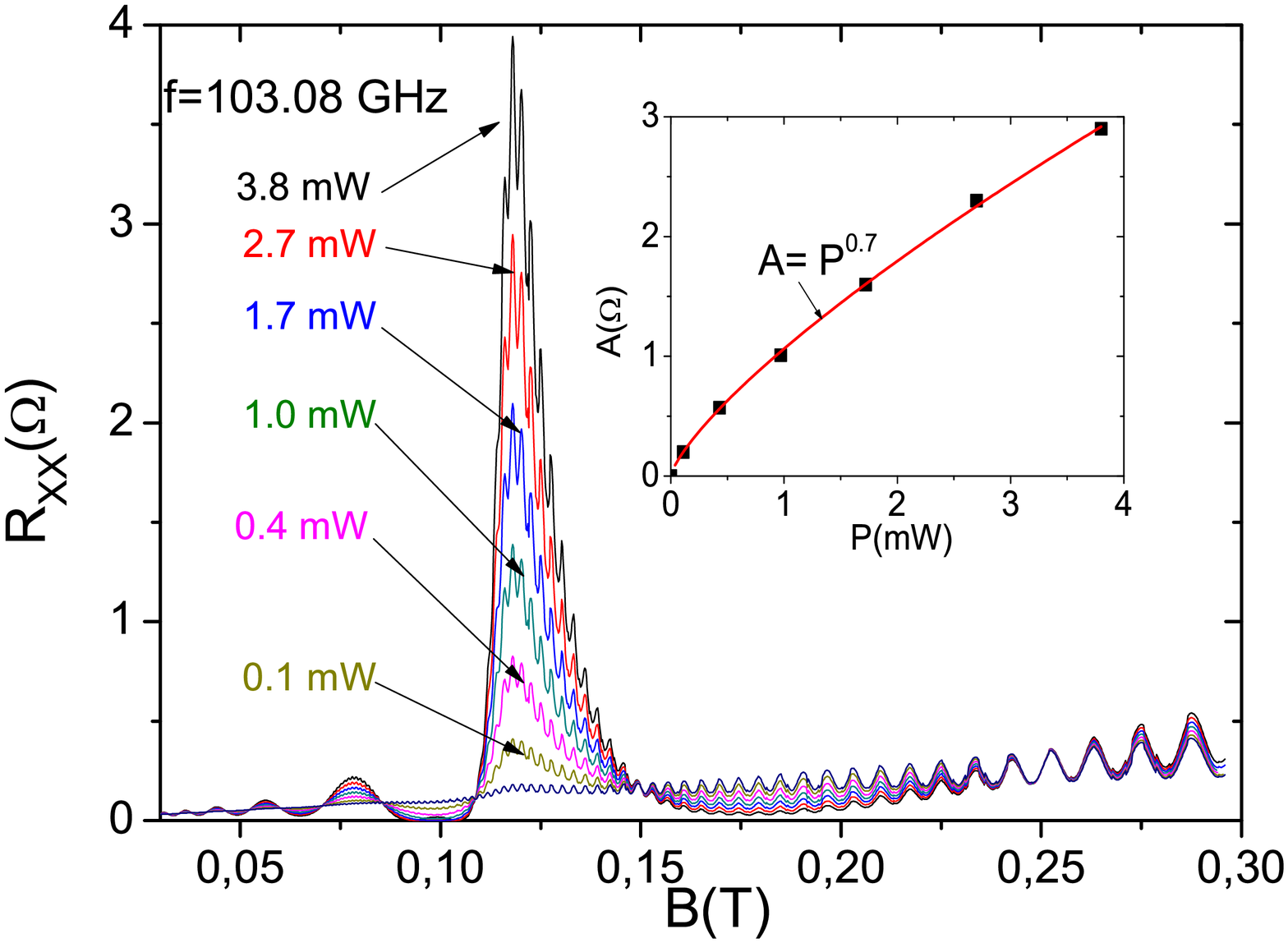}
\caption{Calculated irradiated magnetoresistance vs static magnetic field  for different radiation power  ($P$)
and a  frequency $f=103.08 GHz$.
$P$ varies from 3.8 mW to 0.1 mW, including $P=2.7$, $1.7$, $1.0$ and
$0.4$ mW. For decreasing $P$, the radiation-induced magnetoresistance response
decreases as well, and for $P=0.1 mW$  we obtain a response close
to darkness.
In the inset we present the spike amplitude vs $P$. We fit the data obtaining  a sublinear
$P$-dependence, $R_{xx}\propto P^{\alpha}$ where $\alpha$
is close to $0.5$.}
%($T=1K$ and frequency $f=w/2\pi=143$ GHz.)}
\end{figure}

\subsection{Influence of an in-plane magnetic field}

The effect of an in-plane magnetic field, ($B_{\parallel}$),  on the radiation-induced
resistance oscillations was already studied and published\cite{du}, and
experimental results showed that the main effect was a progressive damping of
the whole resistance response as $B_{\parallel}$ increased.
Subsequent theoretical results\cite{inainplane}, confirmed and
explained the surprising damping in the framework of the
radiation-driven electrons orbits model. Accordingly and as we
said above, in  an irradiated  2DES in the presence of a perpendicular
$B$,
electronic orbits are forced to move back and forth, oscillating
harmonically at the frequency of radiation and with an amplitude
proportional to the radiation electric field.
In their radiation-driven motion,
electrons interact with the lattice ions being damped and producing
acoustic phonons. The presence of $B_{\parallel}$
imposes an extra harmonically oscillating motion in the
$z$-direction enlarging the electrons trajectory in their orbits.
This would increase the interactions with the lattice making the
damping process more intense  and reducing the amplitude of the
orbits oscillations. Therefore, the effect of the presence of  $B_{\parallel}$
 is to increase the disorder in the sample from
the electrons perspective.
%Within our approach is is the interaction
%with acoustic phonons what is going to be mainly affected and increasing
%with $B_{\parallel}$.

The Hamiltonian for electrons confined in a 2D system (x-y plane) by
a potential $V(z)$ and subjected to a total magnetic field
$B_{T}=(B_{x},0,B)$, ($B_{||}=B_{x}$)
%and
%$B_{\bot}=B$)
is given
considering the previous hamiltonian of eq. (1) by:
\begin{equation}
\large
%&&H_{0}=\frac{P_{x}^{2}+P_{y}^{2}}{2m^{*}}+\frac{w_{z}}{2}L_{z}+\frac{1}{2}m^{*}\left[\frac{w_{z}}{2}\right]^{2}
%(x^{2}+y^{2})\nonumber\\
%&&+\frac{P_{z}^{2}}{2m^{*}}+\frac{1}{2}m^{*}w_{x}^{2}
%z^{2}+V(z)+\frac{1}{2}w_{x}z(exB_{z}-2P_{y})\nonumber \\
H=H_{1}-eE_{0}\cos wt X+H_{z}
\end{equation}
%where
%\begin{eqnarray}
%&&H_{xy}=\frac{P_{x}^{2}+P_{y}^{2}}{2m^{*}}+\frac{w_{z}}{2}L_{z}+\frac{1}{2}m^{*}\left[\frac{w_{z}}{2}\right]^{2}
%(x^{2}+y^{2}) \\
%&&H_{z}=\frac{P_{z}^{2}}{2m^{*}}+\frac{1}{2}m^{*}w_{x}^{2}
%z^{2}+V(z)
%\end{eqnarray}
If we consider a parabolic potential for $V_{z}$,
\begin{equation}
V(z)=\frac{1}{2}m^{*}w_{0}^{2} z^{2}
\end{equation}
the Hamiltonian  $H_{z}$  can be written as:
\begin{eqnarray}
H_{z}&=&\frac{P_{z}^{2}}{2m^{*}}+\frac{1}{2}m^{*}(w_{x}^{2}+w_{0}^{2})
z^{2}\nonumber\\
&=&\frac{P_{z}^{2}}{2m^{*}}+\frac{1}{2}m^{*}\Omega^{2} z^{2}
\end{eqnarray}
We have used the
% symmetric gauge for $B_{z}$:
%$\overrightarrow{A_{B_{z}}}=-\frac{1}{2}\overrightarrow{r}\times
%\overrightarrow{B}=(-\frac{y}{2}B_{z},\frac{x}{2}B_{z},0)$, and
 the
Landau gauge for $B_{x}$:
$\overrightarrow{A_{B_{x}}}=(0,-zB_{x},0)$,
%$w_{z}$ is the cyclotron
%frequency of $B_{z}$: $w_{z}=\frac{eB_{z}}{m^{*}}$
and
\begin{equation}
w_{x}=\frac{eB_{x}}{m^{*}}
\end{equation}
%$L_{z}$ is the z-component of
%the electron total angular momentum.
%According to the  experimental
%parameters used\cite{yang,mani2} the hamiltonian term  $
%\frac{1}{2}w_{x}z(exB_{z}-2P_{y})<< H_{xy}+H_{z}$. Then, we can
%discard this term and write:
%\begin{equation}
%$H_{0}\simeq   H_{xy}+H_{z}$
%\end{equation}
%$ H_{xy}$ and $H_{z}$  are the Hamiltonians of two quantum harmonic
%oscillators, the first one is two-dimensional in the $x-y$ plane and
%the last one one-dimensional in the $z$ direction.
The use of a parabolic potential for $V_{z}$ allows that the corresponding time-dependent Schrodinger equation of $H$ can be readily  solved. We
obtain the wave functions of two harmonic oscillators, one
 in the $x$ direction and the other in the $z$-direction:
\begin{equation}
\Psi_{T}(x,z,t)\propto\phi_{N}\left[(x-X-x_{cl}(t)),t\right]
\phi(z)
\end{equation}

 In a semiclassical approach
the electron is subjected simultaneously to two independent harmonic
motions with a trajectory depicted in Fig. 9:
%value of $\gamma \sim 10^{12}s^{-1}$ for GaAs.
As we have indicated above, the
presence of in-plane $B$ alters the electron trajectory in its orbit
increasing the frequency and the number of oscillations in the
$z$-direction. Now the frequency of the $z$-oscillating motion  is
$\Omega>w_{0}$. This makes longer the electron trajectory increasing
the total orbit length and eventually the damping. This increase in
the orbit length is proportionally equivalent to the increase in the
number of oscillations in the $z$-direction. Thus, we introduce the
ratio of frequencies after and before connecting $B_{x}$ as a
correction factor for the damping factor $\gamma$. The final damping
parameter $\gamma_{f}$ is:

\large
\begin{eqnarray}
\gamma_{f}&=&\gamma \times \frac{\Omega}{w_{0}}=\gamma \times
\sqrt{1+\left(\frac{w_{x}}{w_{0}}\right)^{2}}\nonumber\\
&=& \gamma \times
\sqrt{1+\left(\frac{eB_{x}z_{0}^{2}}{\hbar}\right)^{2}}
\end{eqnarray}
\normalsize
where $z_{0}$ is the effective length of the electron
wave function when we consider a parabolic potential for the $z$-confinement\cite{davies,ando},
\begin{equation}
z_{0}=\sqrt{\frac{\hbar}{m^{*}w_{0}}}
\end{equation}
\\

The expression of $\gamma_{f}$  shows that one of the main results
of the presence of $B_{x}$ is an increase
in the damping parameter due to the interaction with
lattice ions. Therefore, the whole resistance response to
radiation, (spike and oscillations) will be
increasingly damped.
The second important effect
%of the increasing
%disorder due to the presence of $B_{x}$
is
reflected in the total quantum scattering rate $1/\tau_{0}$.
According to the Matthiessen rule the total scattering rate can be
expressed as the sum of the different individual scattering sources,
%\begin{equation}
$\frac{1}{\tau_{0}}=\sum_{i}\frac{1}{\tau_{i}}$
%\end{equation}
and obviously one of them is the acoustic phonon scattering rate.
The increase in the phonon scattering
rate ($\gamma=1/\tau_{ac}$) due to $B_{x}$ will eventually affect the total
scattering rate that will increase as well, making it   $B_{x}$-dependent, $\frac{1}{\tau_{0}(B_{x})}$.
Considering the  Matthiessen rule and the ratio of frequencies $\frac{\Omega}{w_{0}}$ affecting
$\gamma$, we can readily obtain the expression,
\large
\begin{eqnarray}
\frac{1}{\tau_{0}(B_{x})}&=&\frac{1}{\tau_{0}}+\gamma\left[\frac{\Omega}{w_{0}}-1\right]\nonumber\\
&=&\frac{1}{\tau_{0}}+\gamma\left[\sqrt{1+\left(\frac{eB_{x}z_{0}^{2}}{\hbar}\right)^{2}}-1\right]\nonumber\\
\end{eqnarray}
\normalsize
%where $\tau_{0}$ is the single particle life time with no magnetic fields present.
And now eq. (40) can be rewritten taking into account the last expression,
%which express that the LL width
\begin{equation}
\large
\Gamma(B,B_{x})\simeq \frac{1}{2\sqrt{0.1}}\sqrt{\frac{\hbar^{2} e}{2\pi m^{*}\tau_{0}(B_{x})}} B^{p}
\end{equation}
Now $\Gamma$ depends on both magnetic fields the perpendicular $B$ and the in-plane $B_{x}$,

\section{Results}

In Fig.1 in the  upper panel, we present a
schematic diagram showing the density of Landau states
for wider and narrow $\Gamma$, (standard and ultraclean
samples respectively).  Therefore, we
can observe the two different regimes; for a standard sample  where $\hbar w_{c}<\Gamma$ and
for a ultraclean one where $\hbar w_{c}>\Gamma$. The broadened Landau states
have been simulated by Lorentzian functions.   The arrow 1 corresponds to an
elastic scattering process (remote charged impurity scattering) and the arrow 2 to an inelastic one
(acoustic phonon scattering).
% Then, according to the model,
%resonance and MIRO's are  displaced at lower $B$, at $w_{c}=\frac{w}{2}$, as in experiments (see Fig. 1).
In the  lower panel of Fig. 1, we present the radiation-driven electron orbit displacement for
ultraclean and standard samples. We observe that the first one is
delayed regarding the latter as being driven by radiation of smaller frequency.

In Fig. 2a,  we present calculated irradiated $R_{xx}$ vs static magnetic field for
a radiation frequency of  $f=103.08 GHz$. The presented results correspond to standard and ultraclean samples and high and low
temperatures. For the
ultraclean sample and $T=0.4 K$, we obtain a strong spike at $w\approx2 w_{c}$ as in
experiments\cite{yanhua,hatke2}. Increasing temperature for the same sample, ($T\simeq 2.5K$)  the spike vanishes but
the radiation-induced oscillations still remain but with lower intensity, as expected. Finally,
for a standard sample, we obtain the usual radiation-induced $R_{xx}$ oscillations and
ZRS. In Fig. 2b we present a blown up of
the lower values of $R_{xx}$ of the upper panel. Thus, we can
contrast the  curve of the ultraclean sample with the
standard, observing a shift of the $R_{xx}$ oscillations to lower magnetic fields.

In our calculations we have used a quantum lifetime at zero
magnetic field $\tau_{0}\sim 2 \times 10^{-11}$ s,\cite{hatke2,chepe,yanhua} for the ultraclean $R_{xx}$ curve, which
gives us, according to eq. (32), a LL width of $\Gamma \sim 5. \times 10^{-4} B^{p} $ eV.
For the standard curve, we have used a shorter quantum
lifetime,\cite{cunkur,studen, wied,gusev}  $\tau_{0}\sim 10^{-12}$ s,
giving a broader $\Gamma$, $\Gamma\simeq 25.10^{-4} B^{p}$ eV.
The quantum life time defines the quality of the sample; the longer this
time, the higher the quality and narrower $\Gamma$.
%This different $\Gamma$, depending on the quality of the sample, will
%gives us different impurity scattering rate and impurity scattering time.
Smaller $\Gamma$, according to eq. (22), will give us a longer impurity
scattering time. Then, the perceived radiation frequency will be
smaller, shifting the $R_{xx}$ oscillations to lower $B$ as observed
in ultraclean samples.
%Eventually reflected in the obtained $R_{xx}$ oscillations that in the case
%of ultraclean samples are shifted to lower $B$ as it happens when
%the sample is illuminated with radiation of  lower
%frequency.
%Then, for
%long $\tau_{0}$ the condition of eq. (20) can be fulfilled, being
%reflected in larger impurity scattering rate (smaller impurity scattering time)
%and eventually smaller perceived radiation frequency.
%In other words, the use of high quality sample (narrow $\Gamma$), makes the
%resistance oscillations to shift to lower magnetic fields.
On the other hand, with a shorter quantum life time (low quality samples or standard),
$\Gamma$ is large and the term inside brackets in eq. (22) tends to $1$. Thus,
we recover the usual positions for radiation-driven $R_{xx}$ oscillations\cite{smet}.

In Fig. 3 we present  calculated irradiated $R_{xx}$ vs magnetic field for different radiation frequencies $f=60.0$, $103.08$,
$137.0$ and $192$ GHz. For all cases we obtain a clear spike at $w\approx2w_{c}$.
We observe that the spike amplitude increases with the frequency and as a result with the
magnetic field. This is because of the presence of the frequency term, $(w_{c}^{2}-\frac{w}{2}^{2})^{2}$
in the denominator of the $R_{xx}$ expression, and then, the higher $w$, the higher the
magnetic field where the resonance takes place.
In the inset of the figure
we observe that the amplitude of the corresponding spike grows with the magnetic field following a  $B^{2}$ law.
This expected from our model\cite{inaw} because according to it,  the radiation-dependent part of the $R_{xx}$
 depends on the
magnetic field as $B^{2}$. Therefore, we predict for the
 amplitude of the  {\it "$w\approx2 w_{c}$ spike"}, a square law dependence on $B^{2}$ that would need
to be experimentally confirmed. The experiments that discovered the $R_{xx}$ spike
showed that the amplitude increases with the frequency. Yet, they do not present
any law or fit showing the  increase rate of the spike amplitude with $B$.

\begin{figure}
\centering \epsfxsize=3.9in \epsfysize=3.9in
\epsffile{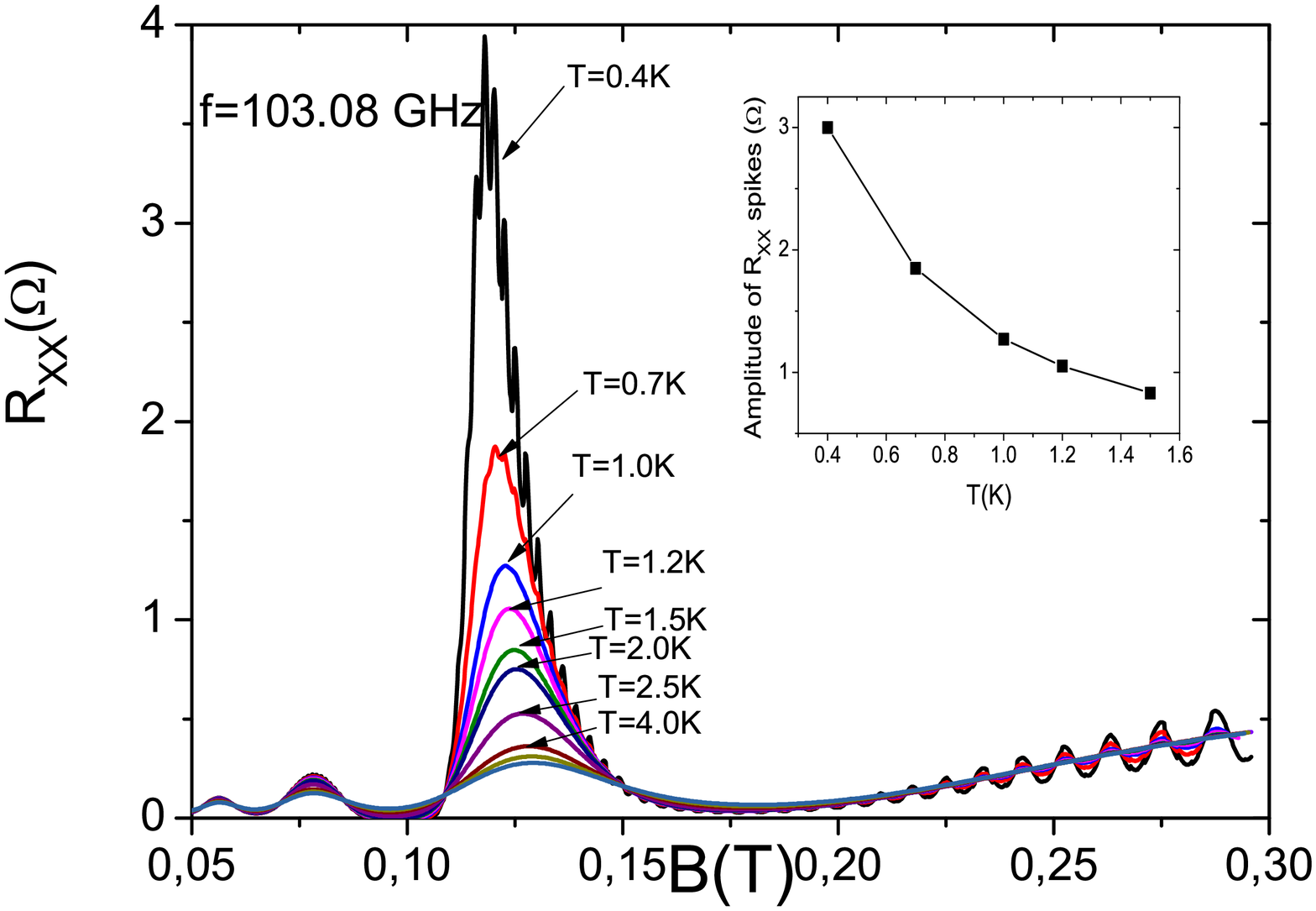}
\caption{Calculated irradiated magnetoresistance vs magnetoresistance for different temperatures
and a frequency $f=103.08 GHz$.
Temperature varies from 0.4 k to 4.0 K.
We observe a clear decrease of the spike and $R_{xx}$ oscillations  for increasing $T$.
The $T$-dependence is explained
with the damping parameter $\gamma$ which represents the interaction of electrons with
acoustic phonons.
In the inset we present the amplitude of the spike vs $T$. We fit the data obtaining a relation $R_{xx}\propto T^{-2}$ (hyperbole), as expected
from the model.}
\end{figure}

\begin{figure}
\centering \epsfxsize=3.5in \epsfysize=5.0in
\epsffile{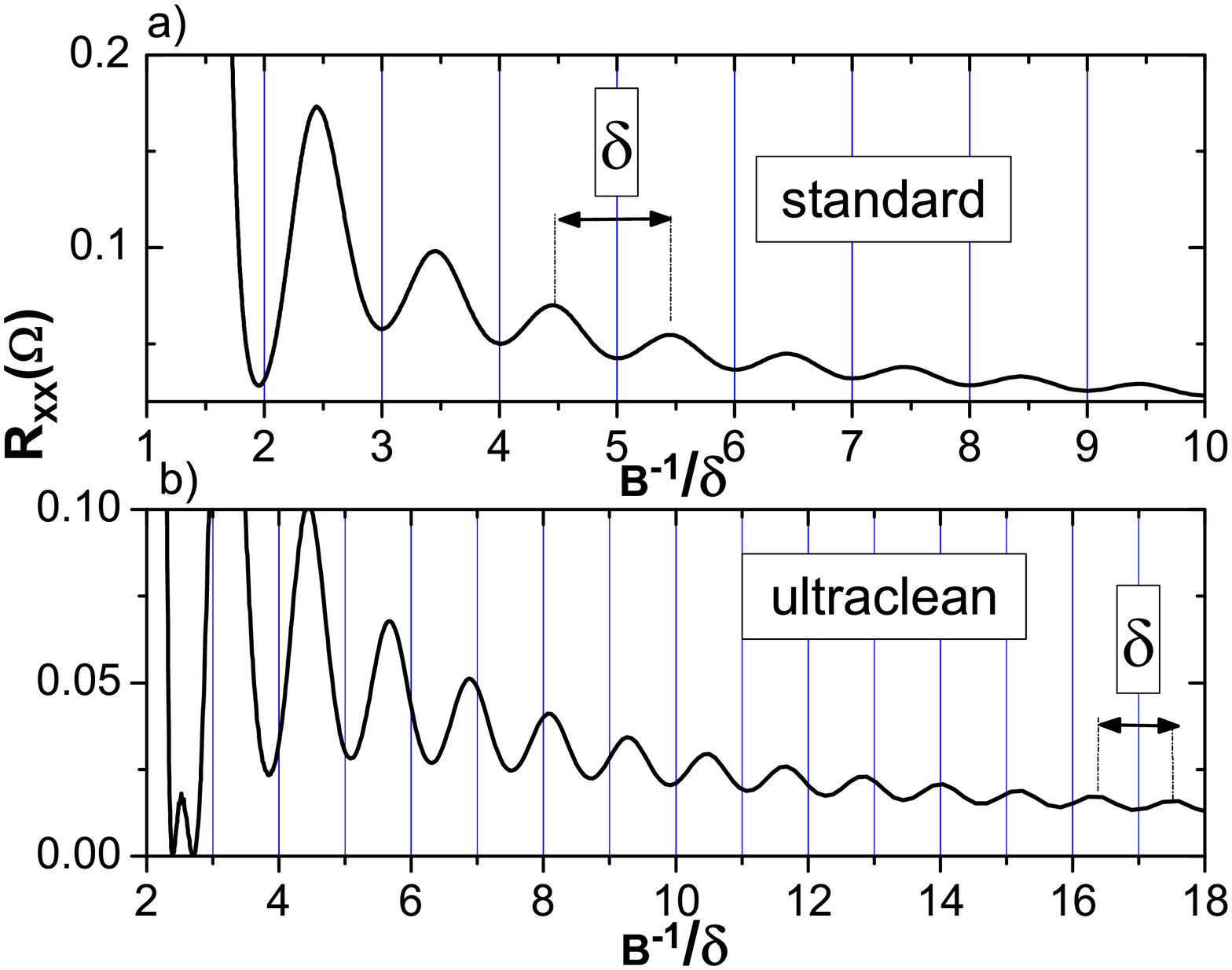}
\caption{Normalized $B^{-1}$ plots of calculated data on
radiation induced resistance oscillations of Fig. 2 for standard sample
(6a) and for ultraclean (6b). $\delta$'s are the oscillatory periods in
$B^{-1}$ for each case.}
\end{figure}

\begin{figure}
\centering \epsfxsize=3.5in \epsfysize=3.0in
\epsffile{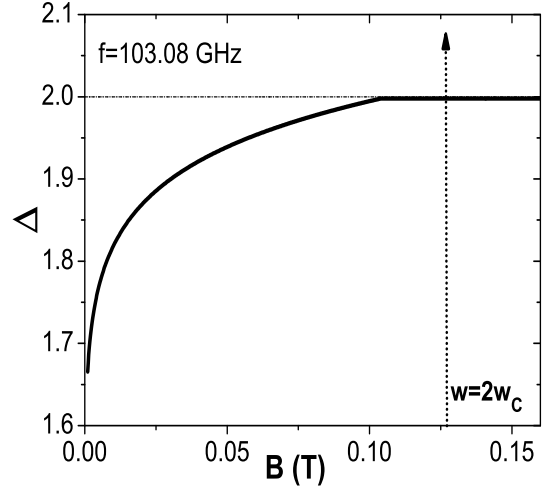}
\caption
{Calculated values for $\Delta$ versus $B$. For $B\simeq 0$ the term $\Delta\rightarrow 1$ because the absolute value  of
the exponents is very large.
Then, as $B$ increases the exponents decrease and $\Delta$
increases till the cut-off value of $2$. This makes the impurity
scattering rate continuously to increase with $B$ from $\Delta\simeq 1$ to $\Delta \simeq 2$, or impurity scattering
time to decrease. This effect is perceived by the electrons as if radiation had
a decreasing frequency. Eventually, the back and forth motion of electrons
in their orbits is performed at a decreasing frequency too.}
\end{figure}

\begin{figure}
\centering \epsfxsize=3.5in \epsfysize=3.5in
\epsffile{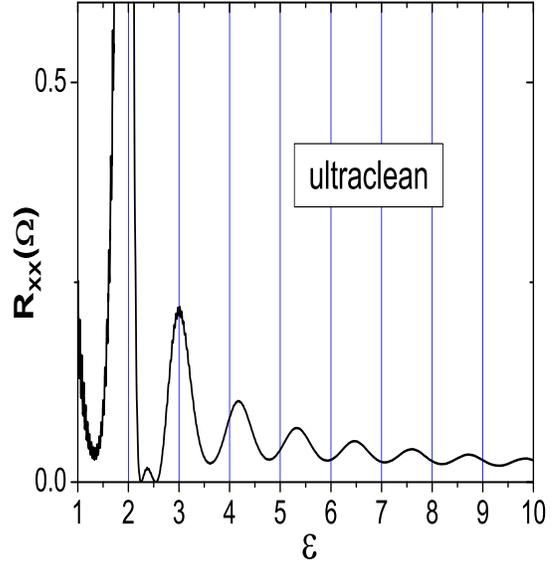}
\caption
{Calculated irradiated $R_{xx}$ vs $\varepsilon$
for a frequency of $f=103.08$ GHz. We observe a qualitatively
similar shift as in experiment\cite{yanhua}. Yet, quantitatively
speaking the calculated shift is larger than the obtained
in the experiment\cite{yanhua}.}
\end{figure}

\begin{figure}
\centering \epsfxsize=3.5in \epsfysize=3.7in
\epsffile{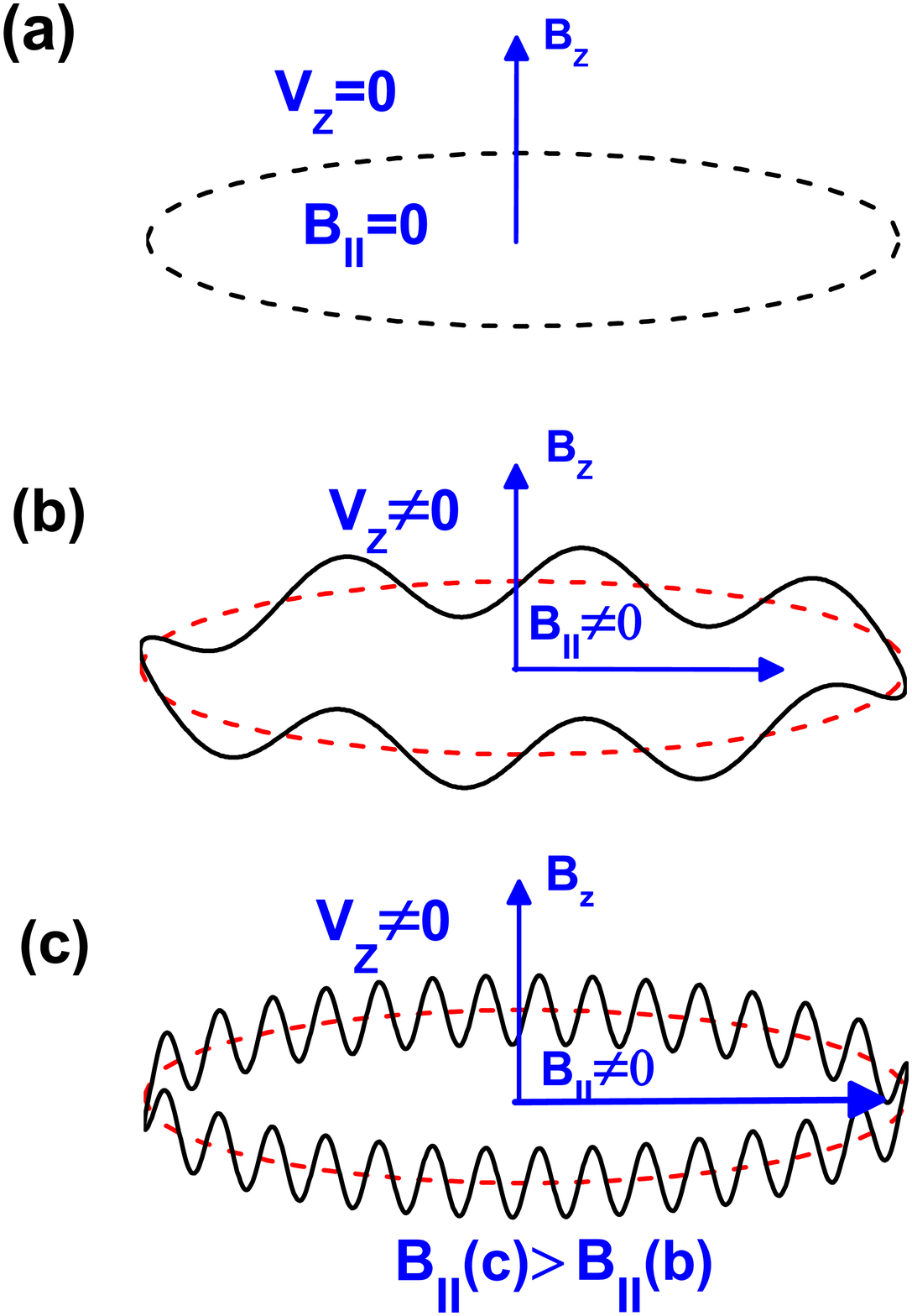}
\caption
{Schematic diagrams showing
the semiclassical description of electron trajectories in 2D systems.
 In the three of them there is
present a perpendicular magnetic field $B_{z}$ that
explains the circular trajectories. a) 2DES ($x-y$ plane) and $B_{z}$. b) 2DES, $B_{z}$, a parabolic
potential $V_{z}$ and an in-plane magnetic field $B_{||}$. c) Same as
 b) but the intensity of $B_{||}$ is bigger.}
\end{figure}

In Fig. 4 we present calculated irradiated $R_{xx}$ vs magnetic field for different radiation power ($P$)
and frequency $f=103.08 GHz$ .
$P$ varies from 3.8 mW to 0.1 mW, including $P=2.7$, $1.7$, $1.0$ and
$0.4$ mW. Decreasing $P$,  the radiation-induced
$R_{xx}$  response, (spike and oscillations),  decreases as well,
  and for $P=0.1 mW$  we obtain nearly the darkness result.
In the inset we present the amplitude of the $R_{xx}$
spike vs $P$. We fit the data obtaining  a sublinear
$P$-dependence,
%\begin{equation}
$R_{xx}\propto P^{\alpha}$
%\end{equation}
where $\alpha$ is close to $0.5$, as
expected,
and explained in terms of our model:
%\begin{equation}
 $E_{0}\propto \sqrt{P}\Rightarrow R_{xx}\propto \sqrt{P}$
%\end{equation}
  and in agreement with
experimental\cite{mani6} and theoretical\cite{ina5} results.
\\
\begin{figure}
\centering \epsfxsize=3.7in \epsfysize=4.5in
\epsffile{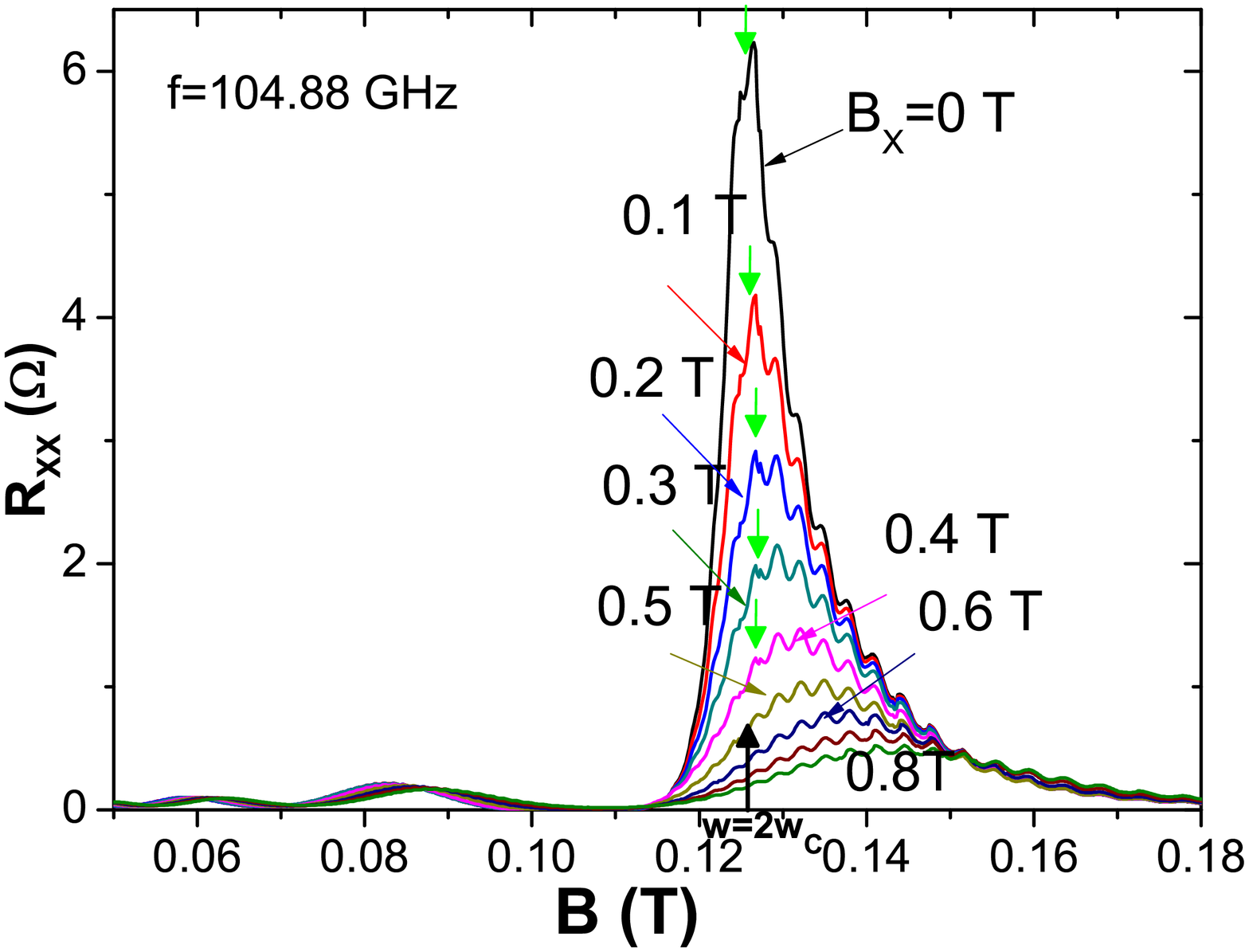}
\caption
{Calculated irradiated $R_{xx}$ versus $B$ for a radiation frequency of
$f=104.88$ GHz and for different values of the in-plane magnetic field  $B_{x}$.
$B_{x}$ goes from 0T till 0.8T in intervals of 0.1T. We first observe a progressive damping
of the resistance spike and resistance oscillations. For approximately $B_{x}=0.6$ T the
spike has been completely removed. We observe as well that starting from $B_{x}=0.5$ and for
larger $B_{x}$ the
main $R_{xx}$ associated with the spike shifts to higher $B$.
However, we observe
that from $B_{x}=0$ T till $B_{x}=5$ T, the $R_{xx}$ spike remains approximately in its position,
 although
its intensity is progressively smaller. The green arrows highlight this effect.}
\end{figure}

\begin{figure}
\centering \epsfxsize=3.7in \epsfysize=5.5in
\epsffile{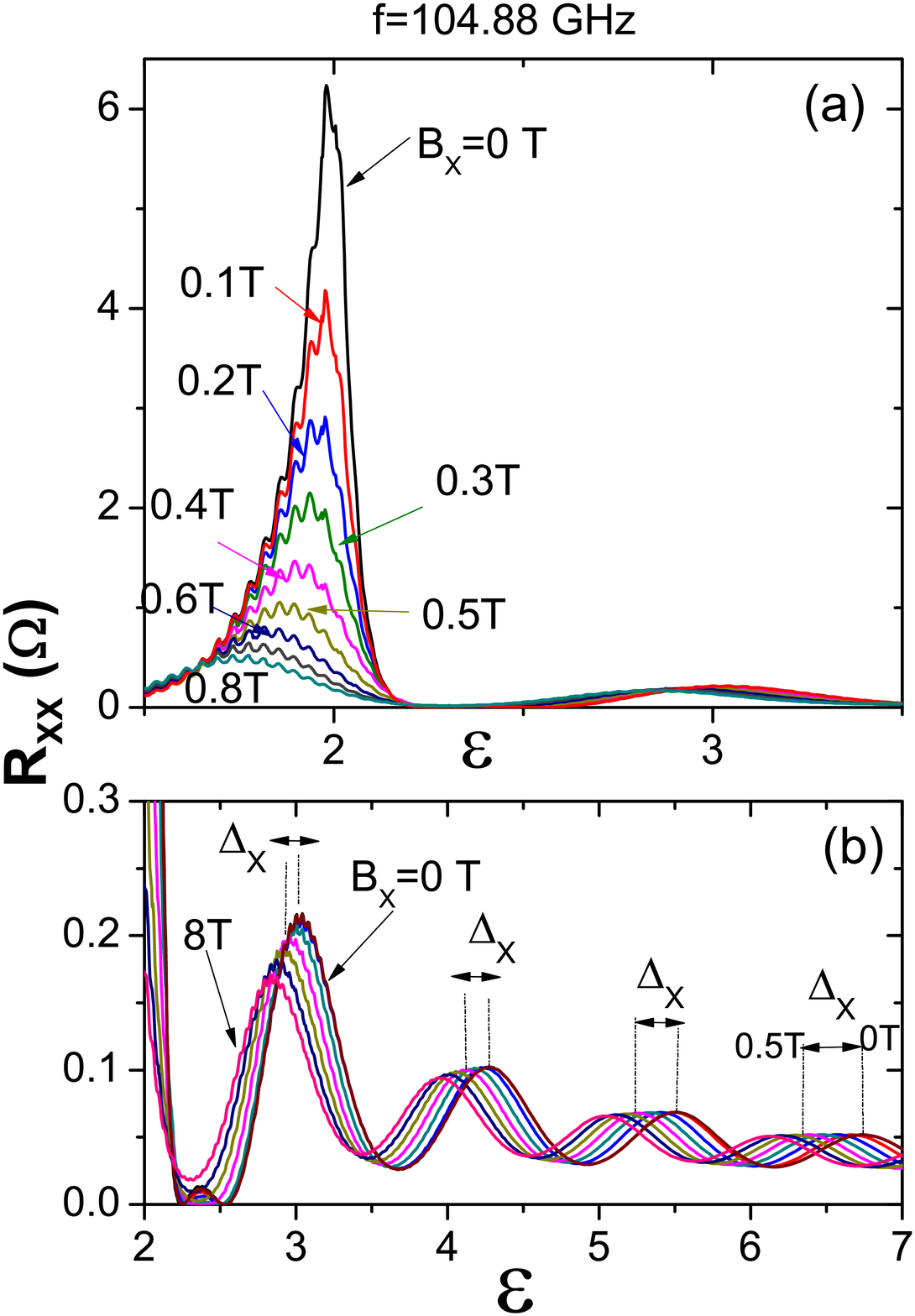}
\caption
{Calculated irradiated $R_{xx}$ vs $\varepsilon$
for a frequency of $f=103.08$ GHz. In a) $\varepsilon$ goes from
1.5 to 3.5. As in Fig. 10 we observe the spike keeping the
same position till $B_{x}=5$ T. Starting from $B_{x}=6$, the main
resistance peak associated with the spike shifts to higher $B$ as
$B_{x}$ increases. In b) $\varepsilon$ goes from
2 to 7. We observe the shift, also to higher $B$, of the other resistance
peaks. The shift is represented by $\Delta_{x}$. Yet, in apparent contradiction, the corresponding spike stands still for the same $B_{x}$  values as observed in a).
}
\end{figure}
In Fig. 5 we present calculated irradiated magnetoresistance vs magnetic field
 and frequency $f=103.08 GHz$, for differente temperatures (T).
$T$ varies from 0.4 to 4.0 K.
We observe a clear decrease of the spike for increasing $T$, till
it is smeared out.
The $T$-dependence, according to the model, is explained
with the damping parameter $\gamma$. $\gamma$ is linear with $T$\cite{ina2,ina3}, and
$R_{xx}\propto \gamma^{-2}$. Thus, an increasing $T$ means
an increasing $\gamma$ and smaller both spike and $R_{xx}$ oscillations.
When the damping is strong enough (higher $T$)
all $R_{xx}$ response collapses giving a final result close to darkness.
The inset  shows the relation $R_{xx}\propto T^{-2}$ (hyperbole), as expected
from the model.
\begin{figure}
\centering \epsfxsize=3.7in \epsfysize=3.5in
\epsffile{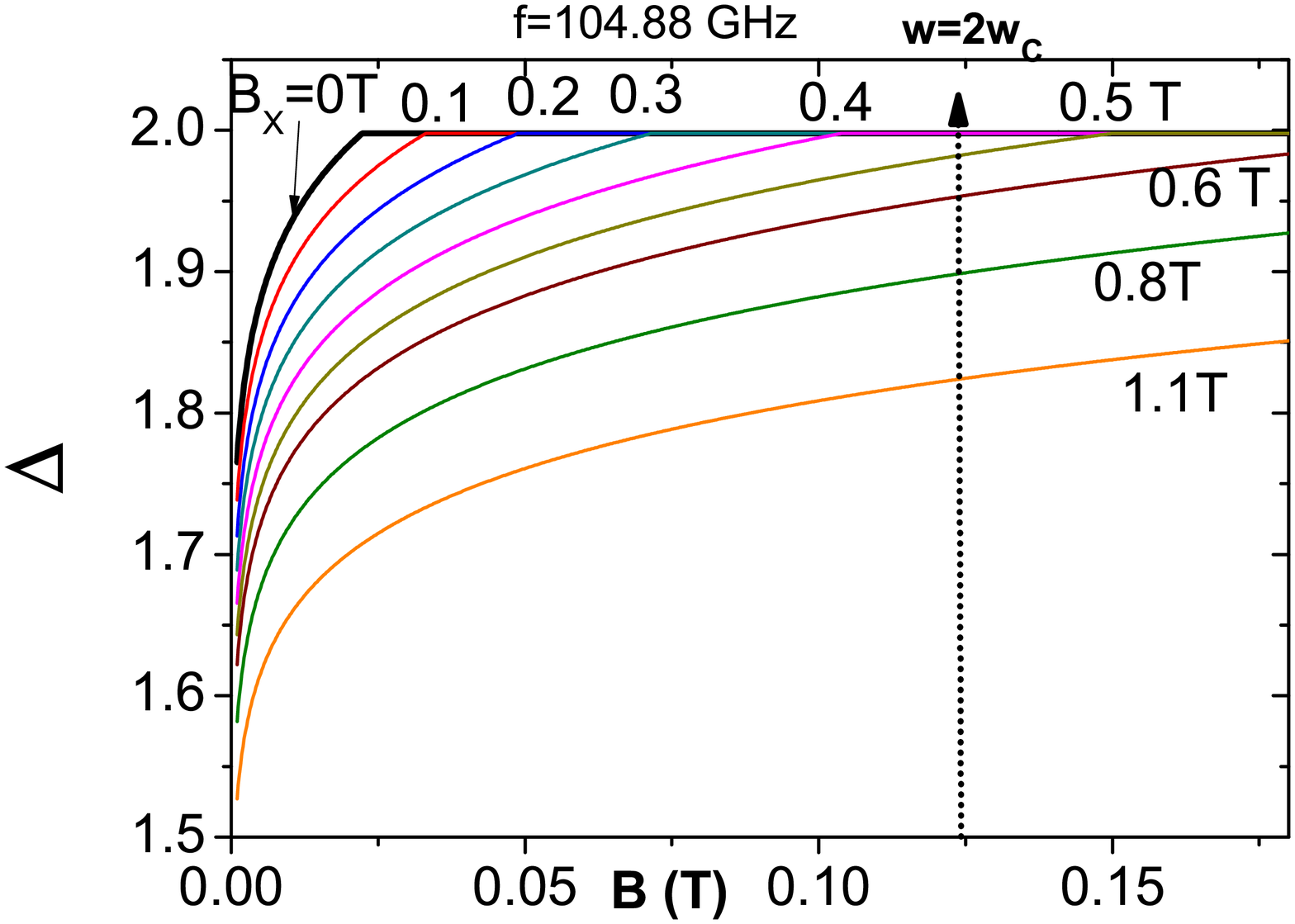}
\caption
{Calculated values of  $\Delta$ term versus $B$ for same values  of $B_{x}$ as in Fig. 10,
and $B_{x}=1.1$ T. The interpretation is similar to Fig. 7.
The presence of an increasing $B_{x}$,  gives rise to larger $\Gamma$ which as a
result makes $\Delta$ to decrease. Then, the impurity scattering time
 decreases continuously being reflected in the perceived radiation
 frequency which decreases too. The  final outcome is that $R_{xx}$ oscillations shift to
 higher $B$ as  $B_{x}$ increases. 
 This is a general behavior for all values of $B_{x}$. 
We can see in the figure that from $B_{x}=0$ T till $B_{x}=5$ the threshold of $\Delta\simeq 2$
is achieved in the sweeping $B$ before the resonance condition of
$w=2w_{c}$ or around it.  Then, when reaching this condition, the resonance effect can take place and
an intense $R_{xx}$ spike will rise in the same position for all of them, }
\end{figure}

In Fig. 6 we present normalized $B^{-1}$ plots of calculated data of
radiation induced $R_{xx}$ oscillations of Fig. 2, ($ f=103.08 GHz$), for standard (6a)
and ultraclean (6b) samples. $\delta$'s are the oscillatory periods in
$B^{-1}$ for each case. In the standard case  we observe
a nearly perfect periodic curve  with respect to $B^{-1}$, as experimentally
observed\cite{mani1}. This is explained with
the linear dependence on $B$ of the charged impurity scattering rate (see eqs. (22) and (23)) when
the brackets term tends to $1$ (standard samples). This implies that the charged impurity scattering time
depends inversely on $B$, giving the periodic curve of Fig. 6a ($R_{xx}\propto A \cos w\tau$).
In the ultraclean case, we observe an oscillatory behavior of $R_{xx}$ versus $B^{-1}$ but
it is not perfectly periodic. Instead, we observe a shift of the $R_{xx}$ oscillations
regarding the vertical lines defined by the period. Peaks and valleys are
increasingly shifted. As $B$ increases, the distance between subsequent
peaks (or valleys) increases.

The origin of this effect
 is in the exponents of eq. (22) which depend on $B$.
The dependence on $B$ of the charged impurity scattering rate comes not only from
the linear term of $W_{0}$ of that equation,  but also from the exponentials in the brackets term
what we call $\Delta$,
\begin{equation}
\Delta=\left( \frac{1+exp\left[\frac{-\pi\Gamma}{\hbar w_{c}}\right]}{1-exp\left[\frac{-\pi\Gamma}{\hbar w_{c}}\right]}\right)
\end{equation}
The variation of $\Delta$ with $B$ is presented in Fig. 7.
For $B$ close to $0$, $\Delta \rightarrow 1$ because the absolute value  of
the exponents is very large.
Then, as $B$ increases the exponents decrease and $\Delta$ continuously
increases till the cut-off value of $2$.
%After that $\Delta$ remains
%in this value as $B$ increases.
This makes the charged impurity
scattering rate continuously to increase with $B$ from $\Delta\simeq 1$ to $\Delta \simeq 2$, (or impurity scattering
time to decrease). This effect is perceived by the electrons as if radiation had
a decreasing frequency. 
%As a result, the back and forth radiation-driven motion of electrons
%in their orbits is performed at a decreasing frequency too.
Eventually all of this is reflected in the obtained $R_{xx}$ where the distance between
$R_{xx}$ peaks (or valleys) gets larger. This is what
explains  the obtained shift in Fig. 6b.

From the experimental standpoint a similar shift can be found
in Fig. 1b of reference [24]. In this Fig. it is represented the
difference between resistivity under radiation minus dark resistivity
versus the parameter $\varepsilon=\frac{w}{w_{c}}$ and obtained an
oscillation shift of $0.25$ of the period. According to the experimental
results\cite{yanhua}, the shift is increasingly larger for decreasing $B$.
To contrast our calculated results on this shift with experiment we present
in Fig. 8 calculated $R_{xx}$ vs $\varepsilon$. We
obtain a qualitative agreement because we
obtain a similar variation of the oscillations shift with $B$.
Yet, quantitatively speaking we obtain a larger value, reaching
$0.5$ of the period. The reason for this quantitative
discrepancy could be explained by the simple
microscopical model used to describe the
dependence of $\Gamma$ on $B$ for ultraclean samples.
On the other hand, the model is able
to explain the existence of the shift, its
variation with $B$ and the connection with
the resistance spike.

In Fig. 9, we present schematic diagrams showing
the semiclassical description of electron trajectories in 2DES (x-y plane),
under a perpendicular magnetic field in the z-direction ($B_{z}=B$). We present three
cases. In a) we have the basic situation of only a  2DES and
$B_{z}$. In the case b) we add to a) a parabolic potential $V_{z}$ and an
in-plane magnetic field $B_{\parallel}$. And finally in the case  c) we have
the same as b) but the intensity of  $B_{\parallel}$ is larger.
In a semiclassical approach
the electron dynamics  under simultaneously these two magnetic fields
can be interpreted as being subjected  to two independent harmonic
motions with trajectories depicted in Fig. 9. When one considers only
 $B_{z}$,  the electron
performs a circular movement in the $x-y$ plane, (see Fig. 9a).  In Fig.
9b, we add a parabolic potential in the $z$ direction and we introduce $B_{\parallel}$, then the
electron trajectory is circular in the plane and at the same time is
oscillating in $z$. In Fig. 9c  $B_{||}$ increases and
the number of oscillations in $z$ direction increases too.

In Fig. 10 we present calculated irradiated $R_{xx}$ versus $B$ for a radiation frequency of
$f=104.88$ GHz and for different values of the in-plane magnetic field  $B_{x}$.
$B_{x}$ goes from $0$ T till $0.8$ T in intervals of $0.1$ T. We first observe a progressive damping
of the $R_{xx}$ spike and oscillations. For approximately $B_{x}=0.6$ T the
spike has been completely removed. We observe as well that for $B_{x}>0.5$,
the main $R_{xx}$ peak associated with the spike, shifts to higher $B$.
As we have previously explained, the main effect of
$B_{x}$ is an increase of disorder in the sample. In our approach this is reflected in
a stronger interaction of electrons with the lattices ions giving rise to a more intense emission of
acoustic phonons. Then, the absorbed radiation energy can be more efficiently released
and the resistance response tends to progressively collapse. This is similar
to the effect of an increasing lattice temperature as previously presented (see Fig. 5). Remarkably we observe
that from $B_{x}=0$ T till $B_{x}=5$ T, the $R_{xx}$ spike remains approximately in its position,
(see green arrows in Fig. 10), although
its intensity is progressively smaller.
%However, for higher $B_{x}$ the resistance oscillations peak corresponding to the spike
%moves to higher $B$ and the spike has been totally smeared out.

In Fig. 11a, we present similar information as in Fig. 10 but this time versus $\varepsilon$.
In Fig. 11b, we present a blown up of 11a for $\varepsilon$ values from 2 to 7. We observe
that meanwhile
the spike does not move when we sweep $B_{x}$ from $B_{x}=0$ T till $B_{x}=5$ T,
the corresponding resistance oscillations present a clear shift to higher $B$. One could
think in principle that such different $B_{x}$ responses could indicate different physical origin. However following our theory, they are totally
connected. And the explanation has to do, as before, with the variation of $\Gamma$
with the magnetic field. The explanation can be obtained from Fig.12.

In Fig. 12 we present the $\Delta$ term versus $B$ for same values  of $B_{x}$ as in Fig. 10,
and adding $B_{x}=1.1$ T. In principle, the interpretation is similar to Fig. 7.
Now the presence of an increasing $B_{x}$,  gives rise to  larger $\Gamma$ which as a 
result makes $\Delta$ to decrease. Then, the impurity scattering time
 decreases continuously being reflected in the perceived radiation
 frequency which decreases too. The  final outcome is that $R_{xx}$ oscillations shift to
 higher $B$ as  $B_{x}$ increases. The effect is similar as having increasingly dirtier samples.
 This is a general behavior for all values of $B_{x}$. The only difference is that 
 for increasing $B_{x}$, $\Delta$ grows slower versus $B$.
Yet, we can see in the figure that from $B_{x}=0$ T till $B_{x}=5$ the threshold of $\Delta\simeq 2$
is achieved in the sweeping $B$ before the resonance condition of
$w=2w_{c}$ or around it.  Then, when reaching this condition, the resonance effect can take place and
an intense $R_{xx}$ spike will rise in the same position for all of them, as experimentally
obtained\cite{yanhua2}. Therefore, we obtain some $B_{x}$ values, ($B_{x}\leq 5$) where constant spike 
position coexist with  a shift of $R_{xx}$ oscillations  to larger $B$.
For the  $B_{x}$ values that reach  $\Delta\simeq 2$ after the resonance
 condition, the spike apparently would rise at higher $B$ but the intense damping effect
done by $B_{x}$ makes it to vanish and can not be observed any longer.
 However, the shift of  $R_{xx}$ oscillations  still can be observed and an example is the shift of
 the main  $R_{xx}$ peak associated with the spike.
 \\

\section{Conclusions}

In summary, we have theoretically studied the  recently discovered intense
radiation-induced magnetoresistance  peak obtained in ultraclean 2DES.
The most remarkable  feature of such a peak is that it shows up
at  $w\approx2w_{c}$ and with an amplitude
of an order of magnitude
larger than the standard radiation-induced magnetoresistance oscillations.
We apply  the radiation-driven electron orbits model
and we calculate the charge impurity
elastic scattering rate, which increases in ultraclean samples, to obtain the novel resonance peak position
(far off resonance).
We obtain the inelastic scattering rate by phonon damping, which decreases in ultraclean samples,
showing that it is responsible of the large peak amplitude.
We present a microscopical model to explain the dependence of the LL width ($\Gamma$) on
the magnetic field for ultraclean samples. We find that this $B$-dependent
variation of $\Gamma$ is essential to explain the experimental shift found in
the resistance oscillations. Accordingly, the shift and the resistance spike are
physically connected.

We have studied also very recent results on the influence of an in-plane magnetic
field on the spike. According to them radiation-driven resistance
oscillations and spike offer different behavior when this magnetic field is increased,
as if they had different physical origin. The same model allows us
to explain such surprising results.
%concluding that both effects
%are sharing the same physics.
We conclude  that the role of the
in-plane magnetic field is mainly to increase the disorder making the
ultraclean sample to behave progressively as a dirtier  sample and
affecting the total scattering rate and the width of the LL. Now, the simultaneous dependence
of this width on $B$ and $B_{x}$ explains the apparent contradiction.
Another important physical effect observed in experiments\cite{yanhua} for the first time in ultraclean
samples is an important decrease in the measured resistance for very low
magnetic fields. This effect is obtained without radiation and is described
in the experiments as a pronounced negative magnetoresistance.
Yet, as the magnetic field increases, the usual values of
standard magnetoresistance are recovered.
This effect, closely related with the quality of the sample,
is the main topic
 of a future work.
 Calculated results are in good agreement with experiments.
These results are remarkable also from a technological standpoint.
For instance in nanophotonics where they could lead to the design and development
of {\it ultrasensitive}
photon detectors in the microwave and terahertz bands where the technology is not
mature yet.

\section{Acknowledgments}

This work is supported by the MCYT (Spain) under grant
MAT2011-24331 and ITN Grant 234970 (EU).

\end{document}